\documentclass[aps,10pt,a4paper,showpacs,jcp,twocolumn,superscriptaddress]{revtex4-1}
\usepackage[utf8]{inputenc}
\usepackage{amssymb}
\usepackage{mathrsfs}
\usepackage{stmaryrd}
\usepackage{graphicx}
\usepackage[usenames, dvipsnames]{color}
\usepackage{epstopdf}
\usepackage{fancyhdr}
\newcommand{\mb}{\mathbf}
\newcommand{\mc}{\mathcal}

\newcommand{\ud}{\mathrm{d}}
\newcommand{\ue}{\mathrm{e}}

\newcommand{\RE}[1]{\textcolor{black}{#1}}

\newcommand{\Nc}{m}

\begin{document}

\title{Phase transitions in single macromolecules: Loop-stretch transition versus loop adsorption transition in
end-grafted polymer chains}

\author{Shuangshuang Zhang}
\affiliation{Department of Physics, Beijing Normal University, Beijing 100875, China}
\affiliation{Department of Basic Courses, Tianjin Sino-German University of Applied Sciences, Tianjin 300350, China}
\affiliation{Graduate School Of Excellence Materials Science in Mainz, Staudingerweg 9, D-55128 Mainz, Germany}
\author{Shuanhu Qi}
\email{qish@uni-mainz.de}
\affiliation{Institut f\"{u}r Physik, Johannes Gutenberg-Universit\"{a}t Mainz, Staudingerweg 9, D-55099 Mainz, Germany}

\author{Leonid I. Klushin}
\affiliation{Department of Physics, American University of Beirut, P. O. Box 11-0236, Beirut 1107 2020, Lebanon}
\affiliation{Institute for Macromolecular Compounds RAS, Bolshoi pr. 31, 199004 St.Petersburg, Russia}

\author{Alexander M. Skvortsov}
\affiliation{Chemical-Pharmaceutical Academy, Professora Popova 14, 197022 St. Petersburg, Russia}

\author{Dadong Yan}
\affiliation{Department of Physics, Beijing Normal University, Beijing 100875, China}

\author{Friederike Schmid}
\affiliation{Institut f\"{u}r Physik, Johannes Gutenberg-Universit\"{a}t Mainz, Staudingerweg 9, D-55099 Mainz, Germany}

\begin{abstract}

We use Brownian dynamics simulations and analytical theory to compare two
prominent types of single molecule transitions. One is
the adsorption transition of a loop (a chain with two ends bound to an attractive
substrate) driven by an
attraction parameter $\varepsilon$, and the other is the loop-stretch transition in a chain
with one end attached to a repulsive substrate,
driven by an external end-force $F$ applied to the free end.  Specifically, we compare the behavior of
the respective order parameters of the transitions, i.e., the mean number of
surface contacts in the case of the adsorption transition, and the mean position of
the chain end in the case of the loop-stretch transition. Close to the
transition points, both the static and the dynamic behavior of chains with different
length $N$ are very well described by a scaling Ansatz with the scaling parameters
$(\varepsilon - \varepsilon^*) N^\phi$  (adsorption transition) and $(F -
F^*) N^\nu$ (loop-stretch transition), respectively, where $\phi$ is the
crossover exponent of the adsorption transition, and $\nu$ the Flory exponent.
We show that both the loop-stretch and the loop adsorption transitions
provide an exceptional opportunity to
construct explicit analytical expressions for the crossover functions
which perfectly describe all simulation results on static properties in the
finite-size scaling regime.  Explicit crossover functions are based on the Ansatz
for the analytical form of the order parameter distributions at the respective transition points.
In contrast to the close similarity in equilibrium static behavior, the dynamic
relaxation at the two
transitions shows qualitative differences, especially in the strongly ordered regimes.
This is attributed to the fact that the surface contact
dynamics in a strongly adsorbed chain is governed by local processes, whereas the end height
relaxation of a strongly stretched chain involves the full
spectrum of Rouse modes.

\end{abstract}

\maketitle
\section{Introduction}

Grafted polymers have attracted great attention in the past few decades due to
their potential applications in surface modification, functional surface
manufacturing, or sensors \cite{advpolymsci.100.31, pps.25.677, pps.28.209,
jpsa.50.3225, jcp.141.204903,ma2015}.  At the level of single molecules, chain
adsorption and chain stretching are among the most prominent fundamental
processes \cite{Zhang_jcp}.  For example, the adsorption of polymers at interfaces can
significantly modify interfacial properties such as the friction
coefficient~\cite{Adsorption}.  On the other hand, tension-induced stretching
is used in modern micromanipulation experiments to characterize the elastic
properties of biological molecules such as DNA or
proteins~\cite{DNA1,DNA2,Protein}. Thus, studying the physics of these two
processes is of basic interest.

In previous work \cite{AdsorbedStretching,klushin97}, two of us have pointed
out a fundamental analogy between the adsorption transition and the
loop-stretch transition of single end-grafted ideal chains.  The adsorption
transition  in the absence of external force ($F=0$)
has been studied intensively for many
decades~\cite{deGennes, Eisenriegler}. It is driven by the competition between
the effective repulsion imposed by a hard substrate, and an additional
attractive interaction with monomers. The effective repulsion is due to the
loss of configurational entropy associated with each contact of the chain with
the impenetrable substrate. If the attractive interaction is weaker than the
entropic repulsion, the tethered chain avoids touching neutral surfaces and
assumes a coiled ``mushroom" configuration. As soon as the attractive energy
exceeds a certain value $\varepsilon^*$ and overcomes the entropic loss, the
chain starts adsorbing onto the substrate. For chains of infinite length,
$\varepsilon=\varepsilon^*$ turns out to be a critical point. The fraction
of monomers in contact with the substrate plays the role of the order parameter
conjugated to the control parameter $\varepsilon$.

\begin{figure}[t]
\centering
\includegraphics[scale=0.5]{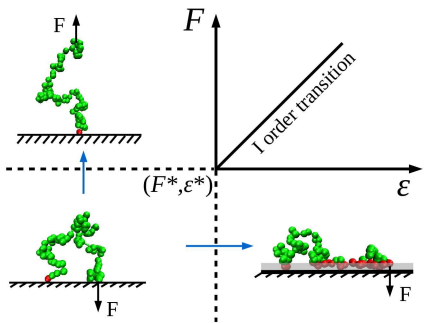}
\caption{
Schematic sketch of phase diagram for chains of infinite length in the
plane of external end force $F$ and adsorption parameter $\varepsilon$, showing
the two transitions investigated in the present work: The loop-stretch
transition (left blue arrow, pointing upwards), and the adsorption transition
(right blue arrow, pointing to the right). The transition points are chosen
such that they are far from the bicritical point at $(F^*, \varepsilon^*)$.
Therefore, the adsorption transition is studied in the presence of a
negative end force. The order parameter characterizing the loop-end transition
is the distance of the chain end from the surface. The order parameter
of the desorption-adsorption transition is the number of chain contacts
with the surface (number of red dots).
} \label{CONF}
\end{figure}

The loop-stretch transition can be observed when a constant force is applied to
the free chain end along the normal direction while the other end is grafted
to the substrate.  Now, the system is characterized by yet another
order parameter conjugated to the end force, $F$: it has the meaning of the
chain stretching and is given by the free end height, $Z$ divided by the chain
contour length \cite{gorbunov93, klushin97, skvortsov10, klushin11, skvortsov12}.
On strongly attractive surfaces the stretching force drives a sharp transition known as mechanical
desorption which turns out to be first order in the infinite chain length limit.
Whereas this transition has been studied quite extensively \cite{BJPS_24_279,JPAMT_48_16FT03,Whittington1,Whittington2},
the situation when the grafting surface is weakly attractive or purely repulsive has received much
less attention. Here, one can still identify a force-driven transition in the infinite
chain limit \cite{gorbunov93} (Fig.\ \ref{CONF}, left arrow).
Depending on the sign of the normal force there appears two types of chain
conformations. Namely, the stretching order parameter tends to zero when the chain end
is pressed onto the substrate (negative force pointing towards the substrate),
or to a finite non-zero limit when the force points away from the substrate. 


At infinite chain length $N \to \infty$, the interplay of the loop-stretch
transition and the adsorption transitions leads to a phase diagram in
the $F-\varepsilon$ plane which is schematically sketched in Fig.\ \ref{CONF}
\cite{klushin97}: Two lines of second order transition, corresponding
to loop-stretch transitions and adsorption transitions, meet at a bicritical
point, which is also the end point of a line of first order transitions
between the zipped and unzipped state on strongly attractive surfaces.
Since a standard discussion of the adsorption transition does not include any end
force, this bicritical point is implicitly identified with the regular critical
adsorption point.

For an ideal continuum Gaussian chain in the presence of a normal end force and
a surface pseudopotential (which could change from attraction to repulsion) one
can establish a formally exact symmetry between the effects of the force and
the surface potential \cite{AdsorbedStretching, klushin97}.  It follows
from this symmetry that  for a given chain length, $N$, the functional
dependence of the average height of the free end, $\langle Z \rangle$, on the
external force, $F$, in the loop-stretch transition (at $\varepsilon=0$)
is the same as the dependence of the average number of adsorbed segments,
$\langle \Nc \rangle$, on the adsorption strength, $\varepsilon$, in the
adsorption transition (at $F=0$). \RE{In the infinite chain limit, it is known that the free energy
for the adsorption of a loop is identical to that of a chain \cite{JPAMG_15_539}.}
In both transitions, finite-size effects are
described by ideal crossover indices (both equal to 1/2) and the same crossover
function. The critical behavior at both transitions can be evaluated within a
Landau expansion and is characterized by mean field critical exponents
\cite{AdsorbedStretching}.

Real chains in good solvent are swollen and the critical behavior in the above
transitions changes. The exponent characterizing the extension of a chain as a
function of chain length $N$ changes from $\nu=1/2$ for random walk to the
Flory exponent $\nu = 0.588$ for self-avoiding walks
\cite{deGennes,Doi_Edwards}. In the vicinity of the adsorption transition, an
independent crossover exponent $\phi$ \cite{Eisenriegler} comes into play,
which characterizes the scaling of the number of contacts with $N$ right at the
critical point. This exponent is not relevant for the loop-stretch
transition. Therefore, one no longer has an exact quantitative correspondence.
At a qualitative level, however, one still expects the loop adsorption
transition and the loop-stretch transition of real grafted chains to share
similar thermodynamic features.

The situation is different when looking at dynamics. Critical singularities in
thermodynamic fluctuations are typically associated with critical slowing down,
and this is also expected here. Beyond that, however, the kinetic behavior of
the two microscopic observables $Z$ and $\Nc$ with respect to $F$ and
$\varepsilon$ should be considerably different. The observable $Z$
characterizing the loop-stretch transition fluctuates
as a result of cooperative motion of monomers involving a full spectrum of Rouse modes
whereas the observable $\Nc$ characterizing the
adsorption-desorption transition fluctuates due to weakly correlated formation
and destruction of monomer contacts separated by large contour length
distances. Thus one expects qualitative differences in the dynamic behavior
at the two transitions.

The purpose of the present paper is to present a systematic comparison of the
loop-stretch transition and the desorption-adsorption transition, both with
respect to static and dynamic behavior. We employ Brownian dynamics (BD)
simulations to investigate the thermodynamic and kinetic behavior of a single
polymer chain grafted onto an impenetrable substrate in the vicinity of the
transitions. To avoid multicritical crossover phenomena, the specific
transition points in the $F-\varepsilon$ plane are chosen such that they are
far from the bicritical point (see Fig.\ \ref{CONF}). Hence, we study the
loop-stretch transition at $\varepsilon = 0$ and the
desorption-adsorption transition at strongly negative $F$, i.e the adsorption of a loop.
The simulations are complemented by
theoretical considerations. Our results confirm and quantify the similarity in
the thermodynamic behavior at the two transitions, and demonstrate the
differences in the relaxation dynamics.

The paper is organized in the following way. In Sec.\ref{sec:model} the model
system and BD scheme are briefly described. Section \ref{sec:result} presents
the analytical theory as well as dynamic simulation results for the
loop-stretch and adsorption transitions. Both static and dynamic properties are
discussed in detail. Section \ref{sec:summary} summaries the present work.
Finally, an Appendix gives some detailed formulations about the BD scheme.

\section{Model and Methods}\label{sec:model}

We consider a polymer chain of $N$ beads connected by Gaussian elastic springs
inside an $L \times L \times H $ box. Throughout this paper, we express all
lengths in unit of the statistical segment length $a$, energies in unit of
$k_\mathrm{B}T$, and times in unit of $\zeta_0 a^2$, where $\zeta_0$ is the
friction coefficient for each bead. We further use $L=4\sqrt{N}$ and $H=N$, and
implement periodic boundary conditions in the $x$ and $y$ directions and
impenetrable boundaries in $z$ direction. One end of the chain is anchored on
the impenetrable surface at $z=0$. To model the stretching process, an external
force $F$, the direction of which is normal to the surface, is applied to the
free end of the chain. To model the adsorption process, an additional
short-range monomer-surface interaction is introduced {\em via} a surface
potential $-\varepsilon \: U_\mathrm{a}(\mathbf{r})$ with $
U_\mathrm{a}(\mathbf{r}) = \min(1,3/2-z)$ (for $z<3/2$).  The strength of
adsorption $\varepsilon$ corresponds to the energy gain if a monomer is in
contact with the substrate. The non-bonded interactions between segments are
described by a coarse grained soft potential, which depends on the local number
density of segments and is evaluated on a grid in the spirit of the
Particle-to-Mesh approach \cite{P3M,PtOM,ma2016} with grid size $b=1$.
The parameter $b$ can be interpreted as a coarse-graining length.
The total Hamiltonian is then given by
\begin{eqnarray}
\label{eq:HH} \mathcal{H}_\mathrm{s} &=&
\frac{3}{2}\sum^{N}_\mathrm{j=2}(\mathbf{R}_\mathrm{j}
  -\mathbf{R}_\mathrm{j-1})^2 + \frac{v}{2} \sum_\alpha \hat{\rho}_\alpha^2
\\ \nonumber && \quad
- F\: Z_1
- \varepsilon \: \sum_{j=1}^{N-1} U_a(Z_j)
\end{eqnarray}
where $\mathbf R_\mathrm{j}$ denotes the location of the $j$th bead,
$Z_\mathrm{j}$ the corresponding z-component (with $Z_N=0$), $\alpha$ runs over
all grid points, and $\hat{\rho}_\alpha$ denotes the local density at the grid
point $\alpha$. Specifically, $\hat{\rho}_\alpha$ is determined from the
particle positions $\mathbf{R}_j$ using the Cloud-in-Cells (CIC) assignment
scheme~\cite{CIC}, with cloud/cell size $b$ and the $z$-components of the
vertices located at $z=0.5 + n$ ($n \in \mathbb{N}_0$). To simplify the
notation for the coordinate of the free end which is one of the main quantities
of interest, in the rest of the paper we drop the subscript: $Z_1\equiv Z$.

The excluded volume parameter $v$ is set to $v=1$ such that the grafted polymer
is in a good (implicit) solvent. The bead positions evolve according to the
equations of over-damped Brownian dynamics, i.e.,
\begin{equation}
\label{eq:BD}
\dot{\mathbf{R}_\mathrm{j}}=-{\partial \mathcal{H}}/{\partial
\mathbf{R}_\mathrm{j}}+ \sqrt{2}\mathbf{f}_\mathrm{r}
\end{equation} where
$\mathbf{f}_\mathrm{r}$ is an uncorrelated Gaussian random force with mean zero
and variance $\langle f_\mathrm{r\alpha}(t)
f_\mathrm{r\beta}(t')\rangle=\delta_{\alpha\beta}\delta(t-t')$
($\alpha,\beta=x,y,z$). The time step for integrating the dynamics equations is
chosen as $\Delta t=0.005$. For details of the BD simulation scheme,
we refer to the Appendix.

The Hamiltonian (\ref{eq:HH}) is used to simulate the stretching process by
setting $\varepsilon=0$ and varying $F$, and to simulate the adsorption process
by setting $F=-2$ and varying $\varepsilon$.  Statistics are performed
for the order parameters of the two processes, i.e., $Z$ for the chain
stretching case and the number of adsorbed segments $\Nc$ for the chain
adsorption case.  Here, $\Nc$ is defined as the number of segments that
experience the surface adsorption potential, i.e., the number of segments with
a distance less than $d=1$ from the surface. The stretching degree $\xi$ and
the fraction of adsorbed segments $\theta$ are defined as $Z/N$ and $\Nc/N$,
respectively.

\section{Results and discussion}\label{sec:result}

\subsection{Characterization of the phase transitions}

We begin with characterizing the thermodynamic properties by examining the
behavior of order parameters and fluctuations. The degree of stretching
$\xi=Z/N$ serves as the order parameter in the loop-stretch problem, where
$Z$ is the distance between the free end and the substrate. Accordingly, the
order parameter in the problem of chain adsorption is the fraction of adsorbed
segments $\theta = \Nc/N$, where $\Nc$ is the number of
adsorbed segments.

The adsorption of a single chain with excluded volume effects (in the force
free case $F=0$) has a long history of exploration.  Monte Carlo (MC) and MD
simulations were used for both lattice and off-lattice models.  Two difficult
problems have to be tackled: The accurate determination of the critical point,
where several methods were proposed and tested, and the evaluation of the
near-critical behavior in the framework of the crossover scaling hypothesis. By
comparison, the loop-stretch transition happens to be much more
straightforward; this may be a reason why it was never studied in detail by
simulations. On the other hand, we will show below that it has the
advantage of admitting an analytical solution.

\begin{figure}[t]
\includegraphics[scale=0.33]{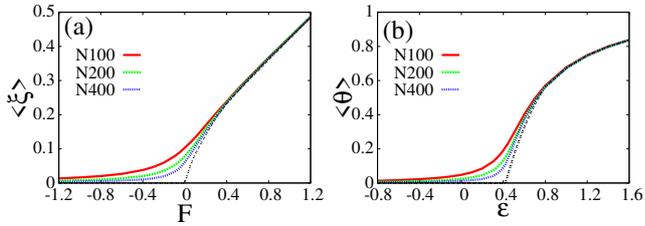}
\caption{(a) Average stretching degree $\langle \xi \rangle$ as a function of
stretching force $F$ for the loop-stretch transition at $\varepsilon=0$
and (b) average fraction of adsorbed segments
$\langle \theta \rangle$ as a function of adsorption strength $\varepsilon$ for
the adsorption transition at $F=-2$ for three chain lengths
$N=100, 200, 400$ as indicated. Black dashed lines show the extrapolation
to the infinite chain limit $N\to\infty$.}
\label{ZC}
\end{figure}

Figs.\ \ref{ZC}(a) and (b) show the ensemble averages $\langle \xi \rangle$
and $\langle \theta \rangle$ as a function of the applied force $F$ and the
adsorption strength $\varepsilon$, respectively.  In both cases, the order
parameter is a continuous function of the control variable, \RE{with an initial slowly increasing part followed by a rapidly increasing region. As the chain length increases, it can be seen that the crossover between these two regions narrows. Albeit hard to achieve numerically, it can be expected that in the infinite chain limit $N\to\infty$, the slope becomes discontinuous at a threshold
point indicating a continuous phase transition in both cases \cite{klushin11}.}

\begin{figure}[b]
\includegraphics[scale=0.34]{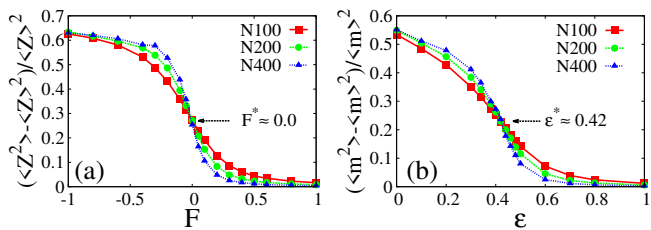}
\caption{Second order Binder cumulants (a) for the loop-stretch transition
at $\varepsilon = 0$, (b) for the adsorption transition
at $F=-2$. The curves for different chain
lengths $N$ intersect at the critical point.} \label{critical_point}
\end{figure}

The critical force value $F^*=0$ for the loop-stretch transition can be
conjectured intuitively. In the infinite chain limit $N\to\infty$ at  $F=0$ the
average stretching order parameter of the mushroom conformation$\langle Z
\rangle/N\sim N^{\nu-1}$ tends to zero. The surface potential is short-ranged
and has no direct relevance as long as it is non-adsorbing. In the same limit
the elasticity of the mushroom vanishes. Hence an arbitrary small (but
$N-$independent) positive end force would produce a nonzero strain. We conclude
that the mushroom state  at $F=0$ is, indeed a critical state. A more formal
rigorous way to locate critical points from the numerical data is the Binder
cumulant method~\cite{Binder_cumulant}. We use the 2nd order cumulants and
determine the critical point as the intersection point of the curves for
$\sigma^2(O)/\langle O \rangle^2$ for different chain lengths $N$,
where $O$ is the relevant order parameter ($O = Z$ and $O=\Nc$, respectively)
and the variance $\sigma^2(O)=\langle O^2 \rangle - \langle O \rangle^2$ quantifies its
fluctuations. The corresponding numerical results, displayed in
Fig.~\ref{critical_point}, give $\varepsilon^*=0.42\pm0.01$ and $F^*=0$. \RE{We notice that the latter result is consistent with the recent studies of self-avioding walks under stretching \cite{BJPS_24_279,JPAMT_48_16FT03}.}

\begin{figure}[t]
\includegraphics[scale=0.34]{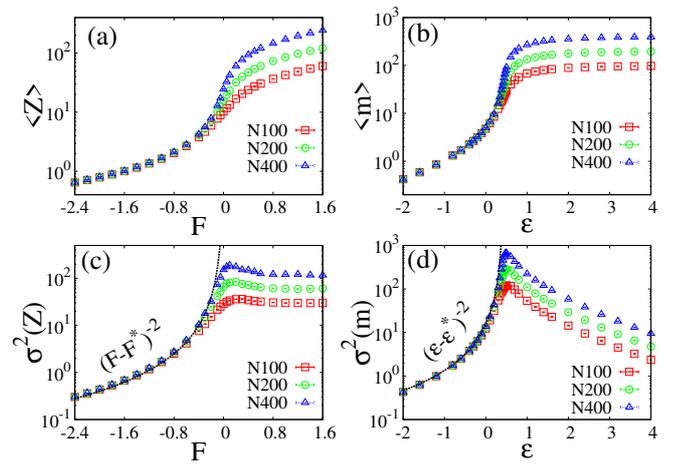}

\caption{Non-normalized order parameters $\langle Z \rangle$ (average distance between chain end and surface)
and $\langle \Nc \rangle$ (number of monomer contacts with the surface) (a,b) and corresponding
fluctuations $\sigma^2$ (c,d) as functions of the conjugate variables $F$
(force applied to the chain end) and $\varepsilon$ (adsorption strength) for
the loop-stretch transition (a,c) and the adsorption transition (b,d) with
chain lengths $N$ as indicated.  Dashed black lines in (c,d) correspond to a
fit of the data for $\sigma^2$ below the transition ($F<F^*$ and
$\varepsilon<\varepsilon^*$, respectively) to an inverse quadratic law.  }
\label{V} \end{figure}

Next we study the behavior of the order parameter and the fluctuations
$\sigma^2$ as a function of the control parameter, i.e., the conjugate
parameter that drives the transition -- the force $F$ in the case of the
loop-stretch transition, and the adsorption strength $\varepsilon$ in the case
of the adsorption transition.  The results are shown and compared to each other
in Fig.\ \ref{V}. Parts (a) and (c) refer to the loop-stretch transition (at
$\varepsilon=0$) and show the chain end position $\langle Z \rangle$ and
$\sigma^2(Z)$, respectively, as a function of the stretching force $F$.  Parts
(c) and (d) refer to the adsorption transition (at $F=-2$) and show the
number of contacts $\Nc$ and $\sigma^2(\Nc)$ as a function of the adsorption
strength $\varepsilon$.  Qualitatively, the curves for the order parameter and
its fluctuations show the same behavior.  Far below the transition, which
corresponds in the loop-stretch case to the loop regime where the
direction of $F$ points to the substrate (i.e., negative $F$ values), and in
the adsorption case to the regime where the adsorption strength $\varepsilon$
is small, the values for the averaged order parameter and its fluctuations
coincide for different chain lengths. As the critical point is approached
($F=F^*$ or $\varepsilon=\varepsilon^*$, respectively), they increase and
become strongly chain length dependent. After crossing the critical point, the
order parameter continues to increase, whereas the fluctuations reach a maximum
and then decrease again slowly.

Table~\ref{tab:notations1} shows the scaling of the quantities under
consideration with respect to the chain length $N$ far below the critical
point, at the critical point, and above the critical point.  Below the critical
point, the quantities do not depend on $N$ and scale with the exponent 0.
Far above the critical point, they increase linearly with the chain length,
i.e., they scale with the exponent 1.  At the critical point, the scaling is
nontrivial and the exponents differ from each other. In the case of the
loop-stretch transition, the scaling of the order parameter $Z$ with chain
length roughly corresponds to the Flory exponent $\nu = 0.59$ \cite{Clisby1}, as one would
expect for a force-free single chain in good solvent, and the scaling exponent
of the fluctuations is twice that number, $2 \nu = 1.18$.  The order parameter
$\Nc$ of the adsorption transition also scales algebraically with the
chain length at the transition point, but with a different exponent
$\phi\approx 0.52$ which is approaching 0.5. This exponent will
discussed further below.
The scaling exponent characterizing the fluctuations of $\Nc$ is
again twice as high, $2 \phi \approx 1.06$.

\begin{table}
  \centering
  \scriptsize
  \begin{tabular}{llll}
    \\[-2mm]
    \hline
    \hline\\[-2mm]
    & {\bf\small below }&\qquad {\bf\small at} &\qquad {\bf\small above}\\
    \hline
    \vspace{1mm}\\[-3mm]
    $\langle Z_1\rangle$
        & $0 \pm0.002$ & $0.59\pm 0.03$ & $1\pm 0.001$ \\
    \vspace{1mm}
    $\langle m\rangle$
        & $0\pm 0.001$ & $0.52\pm0.01$ & $1.005\pm0.003$\\
     \vspace{1mm}
    $\sigma^2(Z_1)$
        &  $0.01\pm 0.01$ & $1.18\pm 0.02$ & $0.98\pm 0.03$\\
     \vspace{1mm}
    $\sigma^2(m)$
        & 0 $\pm 0.002$ & $1.06\pm0.01$ & $1\pm0.01$\\
    \hline
    \hline
  \end{tabular}
\caption{Scaling exponents for the quantities in Fig.\ \ref{V} with chain length $N$ far below the transition ($F=-2.6\ll F^*
=0$ and $\varepsilon=-2.0 \ll \varepsilon^* =0.42$), at the transition
($F= F^*=0.00$ and $\varepsilon= \varepsilon^*=0.42$), and far above the
transition ($F=1.6\gg F^*$ and $\varepsilon=4.0 \gg \varepsilon^*$). }
\label{tab:notations1}
\end{table}

Even though the order parameter and the fluctuations show the same qualitative
behavior at the transition, there are quantitative differences: As we
have seen above, the critical exponents that characterize the scaling of
characteristic quantities with chain length at the transition points are
different.  Moreover, the comparison of Fig.\ \ref{V} (c) and (d) suggests that
the fluctuations of the order parameter decay much more rapidly above the
transition in the case of the adsorption transition than in the case of the
loop-stretch transition. Below the transition, however, the behavior of
$\sigma^2$ at both transitions seems to agree even quantitatively. In
both cases, it can be fitted nicely by an inverse quadratic law,
 $\sigma^2(Z) \sim |F-F^*|^{-2}$ and
$\sigma^2(\Nc) \sim |\varepsilon-\varepsilon^*|^{-2}$, and
it diverges at the critical point in the infinite chain limit
$N \to \infty$.

It is convenient at this point to discuss separately the near-critical
behavior, which will be addressed in the next subsection in the framework
of a finite-size scaling theory, and the behavior far away from the
critical point. The latter is best understood in terms of thermal
``excitations" above the limiting states, which are treated as
``ground states". We first identify the four different ``ground states"
of our system, akin to the cartoons in Fig.\ \ref{CONF} but taken to zero-temperature limit.
\begin{enumerate}
\item[(1)] $F$ fixed, $\; \varepsilon \to -\infty$:
  Mushroom state with no surface contacts ($m=0$)
\item[(2)] $F$ fixed, $\; \varepsilon \to +\infty$:
  Fully adsorbed state, $m=N$
\item[(3)] $F\to -\infty, \; \varepsilon =0$: Loop state with $Z=0$.
\item[(4)] $F\to +\infty, \; \varepsilon =0$: Fully stretched state, $Z = N$.
 This limit cannot be explored in our simulations because of the nature of
 the model, which does not incorporate finite extensibility.
\end{enumerate}
As long as the excitations are independent modes that can be treated within
harmonic approximation, one can apply the equipartition theorem $\langle \Delta
E \rangle = \frac{n}{2}$, where $n$ is the number of independent excitation
modes, and $\Delta E = \mathcal{H}-E_0$ is the deviation of the energy from the
``ground state energy".

In the extreme cases (1) and (2), the only contribution to the energy comes from the surface contacts
($\langle \Delta E \rangle \sim \varepsilon \langle m \rangle$). In case (1)  excitations correspond to isolated contacts bringing positive energy.  Since the repulsive contacts  are dominated by short subchains originating at the grafting point, the effective number of relevant degrees of freedom is
independent of the chain length.  In case (2) excitations are associated with local desorption events that can appear along the whole chain, hence $n$ is proportional to the chain length
$N$.  This yields immediately $\langle m \rangle \sim \:|\varepsilon|^{-1}$ in
the case (1), and $\langle m \rangle \sim \: N-N|\varepsilon|^{-1}$ in the case
(2).  Differentiation with respect to the control parameter $\varepsilon$
results in $\sigma^2(m) \sim \:|\varepsilon|^{-2}$  and $\sigma^2(m) \sim \: N
|\varepsilon|^{-2}$, respectively.

Similarly, in the cases (3,4), the energy is brought by the external end-force: $\langle \Delta E \rangle \sim F \langle Z\rangle$. The excitations are associated with local vertical motions of the
pressed end in the loop case (3), and with the full spectrum of elastic modes
in the stretched case (4). Hence the number of modes that contribute to the
energy is chain length independent in the case (3) and $n \sim N$ in the case
(4). One obtains $\langle Z \rangle \sim \:|F|^{-1}$ and $\sigma^2(Z) \sim
\partial \langle Z \rangle /\partial F \: \sim \:|F|^{-2}$ in the case (3),
and $\langle Z \rangle \sim \: N - N|F|^{-1}$, $\sigma^2(Z) \sim
N\:|F|^{-2}$ in the case (4).


\subsection{Static critical behavior: Theoretical analysis}

\label{theory}

The loop-stretch transition has the advantage of allowing a very detailed
analytical description of the partition function.  We start by summarizing the
known results for various partition functions which are naturally formulated in
the language of lattice models with well-defined discrete configurations. Since
the near-critical behavior of the order parameters is universal, we can
eventually apply the theory to our off-lattice simulations. Firstly recall that
for a free random walk on a lattice, the total number of distinct walks for N
steps is given by

\begin{equation}
Q(N)=\omega^N
\end{equation}
where $\omega$ is the lattice coordination number (e.g., $\omega=6$ for a cubic lattice).
However, for self-avoiding walks of N steps, the total number of walks is
smaller, and asymptotically obeys the law
\begin{equation}
Q(N)=\tilde{\omega}^N N^{\gamma-1}
\end{equation}
where the ``connectivity constant" $\tilde{\omega} < \omega$ depends on the dimensionality
and the lattice type, and the exponent $\gamma$ in the ``enhancement factor"
$N^{\gamma-1}$ depends only on the dimensionality. In three dimensional
lattices, one has $\gamma\approx 7/6$~\cite{deGennes}.

We consider a real chain, modeled as a self-avoiding walk, which is
constrained in half space ($z > 0$) with the one end segment placed at the plane
(i.e., at $z=1$), and define $Q_1$ as the number of walks which starts from the
plane, and $Q_{11}$ as the number of walks which terminates at the plane. It is
reasonable to assume a similar $N$ dependence~\cite{jpamg_11_1833}, i.e.,
\begin{equation}
\label{eq:kappa}
Q_1(N)=\tilde{\omega}^NN^{\gamma_1-1},
\mbox{ and }Q_{11}(N)=\tilde{\omega}^NN^{\gamma_{11}-1}
\end{equation}
The values of $\gamma_1$ and $\gamma_{11}$ have been evaluated both by computer
simulations and theories \cite{ma_7_660,ma_8_946,ma_10_1415,jpamg_10_1927}. \RE{According to the
recent numerical results from Grassberger and Clisby et al, we take $\gamma_1\approx 0.68$ and $\gamma_{11}\approx-0.39$ \cite{jpamg_38_323,Clisby2,Clisby3}.}

%

The form of the probability density distribution for the height of
the free end was conjectured by Fisher ~\cite{jcp_44_616}
\begin{equation}
\label{eq:PP}
P(Z, N)=\frac{A}{Z_0}\Big(\frac{Z}{Z_0}\Big)^{\theta}
\exp\Big[-B\Big(\frac{Z}{Z_0}\Big)^\delta\Big],
\end{equation}
where $Z_0=\langle Z \rangle = b N^\nu$ is the mean end height in the absence of the force,
$b$ is a model-dependent prefactor, and the
coefficients $A$ and $B$
must be chosen such that $P(Z)$ satisfies the normalization conditions $\int_0^\infty P(Z)dZ=1$ and
$\int_0^\infty dZ\: Z\: P(Z)=Z_0$, i.e.,
\begin{equation}
\label{eq:AB}
  A = \frac{\delta}{\Gamma\Big(\frac{1+\theta}{\delta}\Big)}
      \left[\frac{\Gamma\Big(\frac{2+\theta}{\delta}\Big)}
           {\Gamma\Big(\frac{1+\theta}{\delta}\Big)}\right]^{1+\theta},
  \ \
  B = \left[\frac{\Gamma\Big(\frac{2+\theta}{\delta}\Big)}
           {\Gamma\Big(\frac{1+\theta}{\delta}\Big)}\right]^{\delta}.
\end{equation}
The exponents $\delta$ can be derived from
scaling arguments, giving $\delta=\frac{1}{1-\nu}$~\cite{deGennes},
and the exponent $\theta$ follows from the relation
\begin{equation}
\label{eq:PZ1}
Q_{11}=Q_1 \: P(Z=1),
\end{equation}
which gives
\begin{equation}
N^{\gamma_{11}-1}
 =N^{\gamma_1-1}Z_0^{-1-\theta}\sim N^{\gamma_1-1-\nu-\nu\theta},
\end{equation}
and hence~\cite{jpamg_38_323}
\begin{equation}
\label{eq:theta}
\theta=\frac{\gamma_1-\gamma_{11}-\nu}{\nu}\simeq 0.8.
\end{equation}
The coefficients $A$ and $B$ in the distribution function $P(Z,N)$ are
completely defined by two critical indices $\delta$ and $\theta$ with numerical
values of $A\approx 1.18$ and $B\approx 0.503$.  These results hold for
force-free chains that are end-grafted onto a purely repulsive surface ($F=0$
and $\varepsilon=0$).

Structurally, the form of $P(Z,N)$ is a natural generalization of the free end
probability for an ideal Gaussian chain near an impenetrable plane with
``absorbing`` boundary conditions ~\cite{dimarzio_65}.  Although the chain end
distribution is a common object in polymer theory one should also recognize
that in the context of the loop-stretch transition $P(Z,N)$ has the meaning of
the order parameter distribution for a finite system at a critical point.

Based on Eq.~(\ref{eq:PP}), we express the partition function of a
chain subject to an end force $F$ in the force ensemble as
\begin{equation}
Q_F(F,N)=Q_1(N)\!\! \int_0^\infty \!\!\! \ud Z \: P(Z; N) \: \ue^{F\: Z}
  = Q_1(N) \: \Psi_F(F Z_0)
\end{equation}
with
\begin{equation}
  \label{eq:Psiforce}
\Psi_F(x) = A \int_0^\infty \!\! \ud t \: t^\theta
    \exp\Big[-B t^\delta+ x t\Big].
\end{equation}
and $x = F Z_0$.
Hence $Q_F$ has exactly the form assumed in
crossover scaling hypothesis
\begin{equation}
  \label{eq:QFscaling}
  Q_F(F,N)=Q_F(F^*)\: \Psi_F(b F N^\nu),
\end{equation}
where $F^*=0$ defines the critical point, $\Psi_F$ is the crossover
function which contains all information about the static behavior of the
system, and $\nu$ acts as a crossover exponent.
This is  unique case when the crossover function is not just postulated
but can be constructed explicitly.

From the analysis of Eq.(\ref{eq:Psiforce}), one can deduce three regimes
of crossover behavior. In the limit of $x\ll -1$, the second term in the
exponent in Eq.(\ref{eq:Psiforce}) dominates,

\begin{equation}
\Psi_F(x)\sim (-x)^{-\theta-1}.
\end{equation}
In the vicinity of the transition point, at $x \approx 0$, we can expand
$e^{xt}$, which leads to
\begin{equation}
\Psi_F(x)= 1+ x+c_2 \: x^2+\cdots,
\end{equation}
where the coefficient with the linear term equals 1 by normalization conditions, and
\begin{equation}
c_2 =\frac{\Gamma\Big(\frac{1+\theta}{\delta}\Big)\Gamma\Big(\frac{3+\theta}{\delta}\Big)}
           {2\Big(\Gamma\Big(\frac{2+\theta}{\delta}\Big)\Big)^2}
\end{equation}
Numerically, $c_2\approx 0.63$.

\begin{figure}[t]
\centering
\includegraphics[scale=0.45]{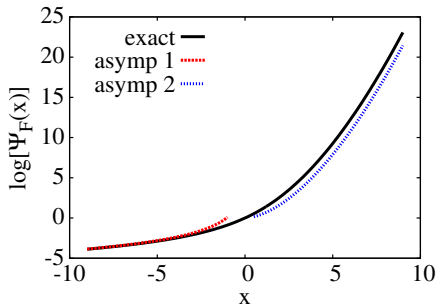}
\caption{ Logarithm of the crossover function of the loop-stretch
transition together with two asymptotic
curves for the loop regime (1) and the stretch regime (2)}\label{psif}
\end{figure}

In the large force limit $x\gg 1$,
the integrand in Eq.(\ref{eq:Psiforce}) becomes very sharp, and we can carry
out a saddle point approximation about the saddle point
$t_c=\big(\frac{x}{B\delta}\big)^{\frac{1}{\delta-1}}$, giving
\begin{equation}
\Psi_F(x)\sim \exp(Cx^{1/\nu})
\end{equation}
where $C=\nu\big(\frac{1-\nu}{B}\big)^{\frac{1-\nu}{\nu}}=0.51$.
Together, we obtain
\begin{equation}
\ln\Psi_F(x)\approx\left\{\begin{array}{ll}
-(1+\theta)\ln(-x), & x\ll -1 \\
x+(c_2-\frac{1}{2})x^2, & |x|\ll 1 \\
C x^{1/\nu}, & x\gg 1
\end{array}\right.
\label{eq:PsiF_2}
\end{equation}
We note that in the pre-transitional regime not only the logarithmic shape
itself but also the numerical prefactor $(1+\theta)$ is predicted to be
model-independent.  The shape of the crossover function $\ln\Psi_F(x)$ is shown
in Fig.\ \ref{psif} together with the two main asymptotics. In the stretching
regime the logarithm of the exact crossover function differs from the
asymptotic branch  by a constant shift which does not affect the behavior of
the observables.

Knowing the crossover function $\Psi_F(x)$, the asymptotic behavior of
the average height can be evaluated according to
$\langle Z\rangle=\frac{\partial\ln\Psi_F(F N^\nu)}{\partial F}$, which
leads to
\begin{equation}
\label{eq:zscaling}
\langle Z\rangle\sim\left\{\begin{array}{ll}
\frac{1}{F^*-F}, & x\ll -1 \\
N^\nu, & |x|\ll 1 \\
N(F-F^*)^{\frac{1-\nu}{\nu}}, & x\gg 1
\end{array}\right.
\end{equation}
Similarly, the asymptotic behavior of the height fluctuations is given
by $\sigma^2(Z)=\langle Z^2\rangle-\langle Z\rangle^2
=\frac{\partial^2\ln\Psi_F (F N^\nu)}{\partial F^2} $, leading to
\begin{equation}
\label{eq:sZscaling}
\sigma^2(Z)\sim\left\{\begin{array}{ll}
\frac{1}{(F-F^*)^2}, & x\ll -1 \\
N^{2\nu}, & |x|\ll 1 \\
N(F-F^*)^{\frac{1-2\nu}{\nu}}, & x\gg 1.
\end{array}\right.
\end{equation}

All theoretical predictions are in good agreement with the simulation data
presented in the previous subsection, Fig.\ \ref{V}(a) and (c).A more
detailed comparison will follow below.

Next we consider the desorption-adsorption transition. A finite-size
scaling hypothesis for the partition function $Q_\varepsilon(\varepsilon, N)$
of the adsorbing chain in a vicinity of the critical point, similar to
Eq.\ (\ref{eq:QFscaling}), was proposed on the basis of the polymer-magnetic
analogy \cite{Eisenriegler, diehl98, MCAdsorption, jpamg_38_323, klushin13}:

\begin{equation}
\label{eq:Qscaling}
 Q_\varepsilon(\varepsilon, N)
 = Q_\varepsilon(\varepsilon^*, N) \:
   \Psi_\varepsilon \big[(\varepsilon-\varepsilon^*)N^\phi\big],
\end{equation}
where
$\phi$ is the crossover exponent for critical adsorption, which has been
discussed extensively in the literature~\cite{Eisenriegler, diehl98,
MCAdsorption, jpamg_38_323, klushin13}.  According to recent extensive lattice
simulations, the value of the crossover exponent $\phi$ is given by $\phi=0.483
\pm 0.003$ \cite{jpamg_38_323, klushin13}.  At chain lengths comparable
to the ones used in the present study, one typically observes higher apparent
exponent close to $\phi=0.5$.  \cite{klushin13}.

The crossover function $\Psi_\varepsilon(x)$ gives the average contact
number $\langle \Nc \rangle = \frac{\partial
\ln \Psi_\varepsilon((\varepsilon - \varepsilon^*) N^\phi)} {\partial
\varepsilon}$. Traditionally, the asymptotic behavior of
$\Psi_\varepsilon(x)$ where $x=(\varepsilon - \varepsilon^*) N^\phi$ is reconstructed from
the expected asymptotic behavior of $\langle \Nc
\rangle$.
\begin{itemize}
\item In the fully developed adsorption regime  $x\gg 1$, the number of adsorbed monomers is proportional to $N$,  $\Nc \sim N$.
Hence, one should have $\ln(\Psi_\varepsilon(x))\sim x^{1/\phi}$ in this
limit.
\item The crossover function is analytic at the transition point $x = 0$, therefore it can be Taylor
expanded for $|x|\ll 1$ giving $\Psi_\varepsilon(x)$ as $\Psi_\varepsilon(x)=1+c_1' x+c_2'x^2+\cdots$
\item For strong enough repulsion $x\ll -1$, very few monomers are contacting the substrate, therefore
$\langle \Nc \rangle$ should be independent of $N$. In this limit,
$\ud \ln(\Psi_\varepsilon(x))/\ud x$ should thus scale like $1/x$, hence
$\ln(\Psi_\varepsilon(x))\sim \ln(-x)$.
\end{itemize}
With these consideration, we summarize the asymptotic expressions of
$\Psi_\varepsilon(x)$ as
\begin{equation}
\ln\Psi_\varepsilon(x)\sim\left\{\begin{array}{ll}
\ln(-x), &  x\ll-1 \\
c_1'x+c_2'x^2, & |x|\ll 1\\
x^{1/\phi},
  & x\gg 1,
\end{array}\right.
\label{eq:PsiE_2}
\end{equation}
and obtain for the average contact number
\begin{equation}
\label{eq:Ncscaling}
\langle \Nc \rangle\sim\left\{\begin{array}{ll}
\frac{1}{\varepsilon^*-\varepsilon}, &   x\ll-1 \\
N^\phi, & |x|\ll 1 \\
N(\varepsilon-\varepsilon^*)^{\frac{1-\phi}{\phi}},
& x \gg 1.
\end{array}\right.
\end{equation}
The fluctuations of $\Nc$ are calculated as $\sigma^2(\Nc)
=\langle \Nc^2\rangle-\langle \Nc\rangle^2=
\frac{\partial^2\ln\Psi_\varepsilon ((\varepsilon - \varepsilon^*) N^\phi)}
{\partial\varepsilon^2}$, yielding
\begin{equation}
\label{eq:sNcscaling}
\sigma^2(\Nc)\sim\left\{\begin{array}{ll}
\frac{1}{(\varepsilon^*-\varepsilon)^2}, & x\ll -1 \\
N^{2\phi}, & |x|\ll 1\\
N(\varepsilon-\varepsilon_c)^{\frac{1-2\phi}{\phi}},
& x\gg 1,
\end{array}\right.
\end{equation}
which is in qualitative agreement with Fig.\ \ref{V} (c) and (d) if
one assumes $\phi=0.52$.

Earlier we have seen that this type of asymptotic behavior of the
crossover function follows naturally from its connection to a specific shape
of the probability density for the order parameter at the transition point. Thus we propose a
Fisher-type conjecture, see Eq.~(\ref{eq:PP}) for the distribution of the
number of contacts at the critical adsorption point.
\begin{equation}
\label{eq:Pcont}
P(\Nc,N)=\frac{A'}{\Nc_0}\Big(\frac{\Nc}{\Nc_0}\Big)^{\theta'}
\exp\Big[-B'\Big(\frac{\Nc}{\Nc_0}\Big)^\frac{1}{1-\phi}\Big].
\end{equation}
Here $\Nc_0=\langle \Nc \rangle =b' N^\phi$ is the average number of
contacts at the critical point, $b'$ is a model-dependent prefactor,
index $\delta'$ is related to the crossover
exponent as $\delta'=(1-\phi)^{-1}$,  and $\theta'$ is yet another exponent
describing the probabilities of chain conformations with very few contacts at
critical conditions $\varepsilon=\varepsilon^*$. The coefficients $A'$ and $B'$
are determined by normalization and expressed in terms of the exponents
$\theta'$ and $\delta'$ according to Eq.(\ref{eq:AB}).

\begin{figure}[t]
\hspace*{-3mm}
\includegraphics[scale=0.35]{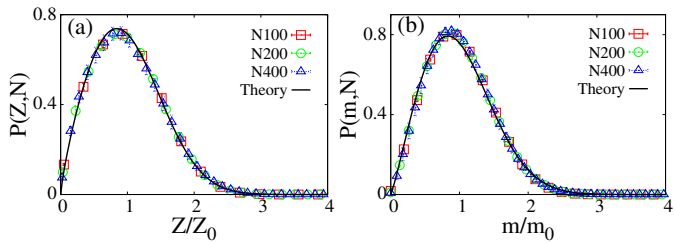}
\caption{Rescaled order parameter distribution at the critical point
for the loop-stretch transition (a) (at $\varepsilon = 0$) and the
adsorption transition (b) (at $F = - 2$) for different chain lengths
as indicated, compared to the predictions of the theoretical conjectures
Eq.\ (\ref{eq:PP}) with exponents $\theta = 0.8$ and $\nu = 0.58$ in (a)
and Eq.\ (\ref{eq:Pcont}) with exponents $\theta' = 1.1$ and
$\phi = 0.52$ in (b).
}
\label{op_distributions}
\end{figure}

Following the  discussion around Eqs.(\ref{eq:PZ1})-(\ref{eq:theta}) we
write
\begin{equation}
\label{eq:Pm1}
Q_{11}=Q_{11c} \: P(m=1),
\end{equation}
where
\begin{equation}
Q_{11c}(N)=\tilde{\omega}^NN^{\gamma_{11c}-1}
\end{equation}
is the partition function of the loop at the critical adsorption point, and
$\gamma_{11c}$ is the corresponding surface exponent. Numerically, $\gamma_{11c}\approx 0.707$ \cite{jpamg_38_323}.
Equation (\ref{eq:Pm1}) leads to
\begin{equation}
\label{eq:thetap}
\theta'=\frac{\gamma_{11c}-\gamma_{11}-\phi}{\phi}\simeq \RE{1.1}
\end{equation}
in full analogy to  Eq. (\ref{eq:theta}).
We test the conjectures (\ref{eq:PP}) and (\ref{eq:Pcont}) by
comparing them with the corresponding order parameter distributions
at the respective critical conditions as obtained from the simulations. The results
are shown in Fig.\ \ref{op_distributions}. If plotted as a function of the relevant
scaling variable, $Z/Z_0$, and $m/m_0$, the theoretical predictions
as well as the distribution
functions for different chain lengths obtained from simulations collapse almost perfectly.
Note that theoretical conjectures contain no fitting parameters at all.

From Eq.\ (\ref{eq:Pcont}), we can construct the adsorption crossover
function $\Psi_\varepsilon(x)$ according to Eq. (\ref{eq:Psiforce}) with proper
replacement of the parameters, leading to a similar form:
\begin{equation}
  \label{eq:PsiE}
\Psi_\varepsilon(x) = A' \int_0^\infty \!\! \ud t \: t^{\theta'}
    \exp\Big[-B' t^{\delta'}+ x t\Big]
\end{equation}
where  $x=(\varepsilon-\varepsilon^*)\Nc_0 = b'(\varepsilon-\varepsilon^*)N^\phi$.
According to this equation, the differences
between the equilibrium near-critical behavior at the loop-stretch and the loop adsorption
transitions for large chain lengths, $N \to\infty$, can thus be traced back to the differences in the
pair of exponents, of which one ($\theta,\theta'$) shows up directly in
the amplitude of the pre-transitional branch.  The post-transitional growth of the order
parameter is directly affected by the crossover exponents,$\nu$ and $\phi$, respectively, and indirectly
(through the coefficients $B$) by the value of the $\theta$ exponent. Interestingly,
the crossover exponents are rather close numerically while the other pair differ more considerably.
We shall see, though, that the width of the near-critical scaling region in
terms of the respective control parameters ($F$ and $\varepsilon$) is dramatically
smaller for the loop adsorption.

\subsection{Simulation results in the finite-size scaling framework}

\begin{figure}[t]
\hspace*{-3mm}
\includegraphics[scale=0.35]{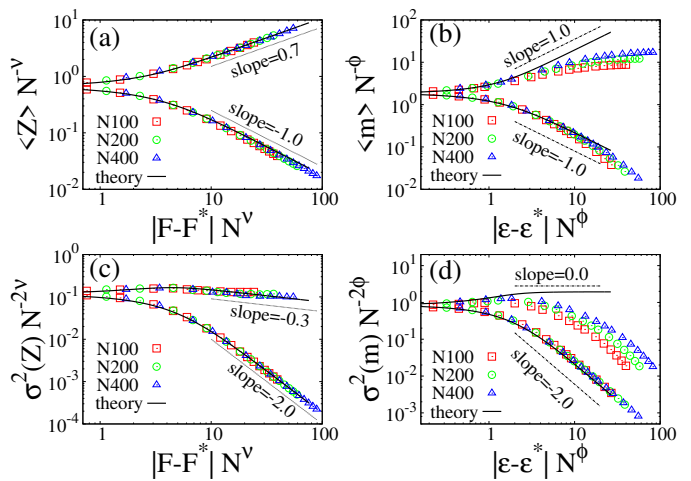}
\caption{Same data as Fig.\ \protect\ref{V}, rescaled as suggested by Eqs.\
(\ref{eq:Fscaling}) (a,c) and (\ref{eq:Escaling}) (b,d).  Lower branch
indicates the regime below the critical point ($F<F^*=0$ and $\varepsilon <
\varepsilon^*=0.42$, respectively), and upper branch the regime above the
critical point ($F>F^*$ and $\varepsilon > \varepsilon^*$, respectively).
Black solid lines represent the predictions of the explicit crossover
functions. Dashed lines show asymptotically expected slopes far from the
transition, if one assumes that the order parameters and their fluctuations
scale like $N^0$ in the lower branch, and like $N^1$ in the upper branch.  }
\label{critical_expos}
\end{figure}

According to the analysis developed in the last subsection for the loop-stretch
transition, the results for $\langle Z \rangle$ and $\sigma^2(Z)$ can also be
written in a scaling form,
\begin{eqnarray}
\label{eq:Fscaling}
\langle Z\rangle &=& N^\nu \hat{f}_{_{Z}}((F-F^*) N^\nu),
\\ \nonumber
\sigma^2(Z)&=& N^{2 \nu} \hat{f}_{\sigma^2_{Z}} ((F-F^*) N^\nu),
\end{eqnarray}
where the scaling functions $\hat{f}_{_{Z}}(x) \propto \ud \ln\Psi_F(x)/\ud x$
and $\hat{f}_{\sigma^2_{Z}}(x) \propto \ud^2 \ln\Psi_F(x)/\ud x^2$ show
asymptotic scaling behavior $\hat{f}(x) \sim (\pm x)^{\alpha_{\pm}}$ for large
$\pm x \gg 1$ with exponents $\alpha_- = -1$ and $\alpha_+ = 1/\nu - 1 \approx
0.7$ for $\hat{f}_{_{Z}}$, and $\alpha_-=-2$, $\alpha_+=1/\nu -2 \approx -0.3$
for $\hat{f}_{\sigma^2_{Z}}$.  This implies that the curves of $\langle Z
\rangle$ and $\sigma^2(Z)$ for different chain lengths $N$ should collapse, if
we plot $\langle Z \rangle N^{-\nu}$ and $\sigma^2(Z) N^{-2 \nu}$ as a function
of the scaling variable $x \sim (F-F^*) N^\nu$. The rescaled data are displayed
in Fig.~\ref{critical_expos}(a) and (c) together with theoretical curves
evaluated from the explicit crossover function of Eq.\ (\ref{eq:Psiforce}) with the
numerical value $b=0.66$ taken from simulations.
Indeed, the data collapse nicely, and are in perfect agreement with the
theoretical curves.

In the same vein, the data of Fig.\ \ref{V} (c,d) should collapse onto one
curve if one plots $\langle \Nc\rangle N^{-\phi}$ and $\sigma^2(\Nc)
N^{-2\phi}$.  as a function of the scaling variable
$x=(\epsilon-\epsilon^*)N^\phi$.  More specifically, the results
(\ref{eq:Ncscaling}) and (\ref{eq:sNcscaling}) can be rewritten in a scaling
form
\begin{eqnarray}
\label{eq:Escaling}
\langle \Nc \rangle &=& N^\phi
  \hat{f}_{_{\Nc}}((\varepsilon-\varepsilon^*) N^\phi),
\\ \nonumber
\sigma^2(\Nc)&=& N^{2 \phi}
   \hat{f}_{\sigma^2_{\Nc}} ((\varepsilon - \varepsilon^*) N^\phi),
\end{eqnarray}
where the scaling functions $\hat{f}_{_{\Nc}}(x) \propto \ud
\ln\Psi_\varepsilon(x)/\ud x$ and $ \hat{f}_{\sigma^2_{\Nc}}(x)  \propto
\ud^2 \ln\Psi_\varepsilon(x)/\ud x^2$ show asymptotic scaling behavior
$\hat{f}(x) \sim (\pm x)^{\alpha_{\pm}}$ for large $\pm x \gg 1$ with $\alpha_-
= -1$ and $\alpha_+ = 1/\phi - 1 \approx 1$ for $\hat{f}_{_{\Nc}}$, and
$\alpha_-=-2$, $\alpha_+=1/\phi -2 \approx 0.$ for $\hat{f}_{\sigma^2_{\Nc}}$.
The rescaled data are shown in Fig.\ \ref{critical_expos} (b) and (d),
together with the theoretical curve derived from Eq.\ (\ref{eq:PsiE}) with the
numerical value $b'=1.85$ taken from simulations.  Even
though $\phi$ has been adjusted to optimize the collapse, the data collapse is
not nearly as good as in the case of the loop-stretch transition.  The best
collapse is obtained for $\phi = 0.52$, which is consistent with the scaling
suggested by Fig.\ \ref{op_distributions} and close to the
literature value, $\phi \approx 0.483$ \cite{jpamg_38_323,klushin13}.
In the near-critical regime, the theoretical curves obtained from
the conjectured explicit form of the crossover function,
Eq.\ (\ref{eq:PsiE}), compare well with the simulation data.

However, at some distance from the transition, the curves for different $N$
do not collapse and the asymptotic behavior of the curves does not reproduce
the theoretically predicted slopes for the chain lengths under consideration.
The main reason for these deviations is that the near-critical behavior at
adsorption is apparently restricted to a rather narrow range of the adsorption
parameter $\varepsilon$  near $\varepsilon^*$.  For stronger adsorption,
saturation effects start suppressing the power-law growth of the contact
numbers which leads to a corresponding decrease in their fluctuations. This
phenomenon falls outside the crossover Ansatz; deviations, however, cannot be seen
at the level of the probability density of Fig. \ref{op_distributions}
as they pertain to the far-off tail at large values of the order parameter.

\subsection{Dynamical behavior}

\begin{figure}[t]
  \includegraphics[scale=0.35]{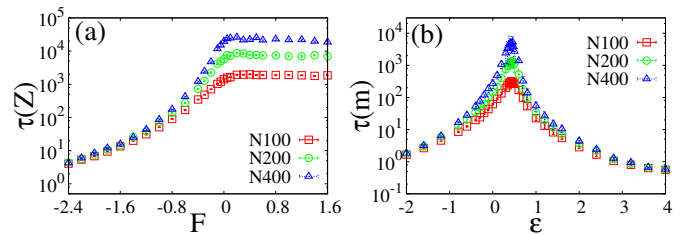}
\caption{Characteristic relaxation times of the order parameter $Z$
(chain end position), $\tau(Z)$, and $\Nc$ (number of monomer contacts with the surface), $\tau(\Nc)$, as
 functions of the respective control parameters $F$ (force applied to the chain
end) and $\varepsilon$ (adsorption strength) (a) at the loop-stretch transition
($\varepsilon = 0$), and (b) at the loop adsorption transition ($F=-2$) and
different chain lengths as indicated.}
\label{relaxation_time}
\end{figure}

After studying the static critical behavior, we turn our attention to the
dynamic behavior of the order parameters. Relaxation of the free end for a
chain grafted to a repulsive surface  was never studied in the context of the
loop-stretch transition. As for the relaxation of the number of contacts, it
was evaluated only at the critical point of adsorption in  \cite{MCAdsorption}.
We calculate the autocorrelation function of the two order parameters, i.e.,
$C_{Z}(t)=\frac{\langle Z(t)Z(0) \rangle-\langle Z \rangle ^2}{\sigma^2(Z)}$
and $C_{\Nc}(t)=\frac{\langle \Nc(t)\Nc(0) \rangle-\langle \Nc \rangle
^2}{\sigma^2(\Nc)}$, from which the characteristic relaxation times $\tau(Z) =
\int_0^\infty \!\! \mathrm{d} t \: C_{Z}(t)$ and $\tau(\Nc) = \int_0^\infty
\!\! \mathrm{d} t \: C_{\Nc}(t)$ can be evaluated. The results are presented in
Fig.~\ref{relaxation_time}.

Below the phase transition, for negative forces or non-adsorbing surfaces,
respectively, the relaxation times show a similar behavior at the two
transitions. They are almost independent of chain length and increase as the
critical point is approached. Close to the critical point, they become strongly
chain length dependent and develop a maximum, which would diverge in the limit
$N \to \infty$. (Note that the maximum in the case of the loop-stretch
transition is much less prominent but still exists).  Above the transition,
however, the curves in Figs.\ \ref{relaxation_time}(a) and (b) are markedly
different.  In stretched chains, the relaxation times remain large and still
depend strongly on the chain length, whereas in adsorbed chains far above the
adsorption point, they drop down again and become roughly chain length
independent.  This is also apparent from Table II, which shows the scaling of
$\tau$ with the chain length $\tau \sim N^\alpha$, far from the transition and
at the transition. Far below the transition, one has $\alpha = 0$ both for the
loop-stretch and the adsorption transition. At the transition, the scaling
exponent $\alpha$ is nontrivial and close to $2$ in both cases. Far above the
transition, one recovers $\alpha = 0$ in the adsorption case, whereas $\alpha$
settles at $\alpha=2$ in the loop-stretch case.

\begin{table}
  \centering
  \scriptsize
  \begin{tabular}{llll}
    \\[-2mm]
    \hline
    \hline\\[-2mm]
    & {\bf\small below }&\qquad {\bf\small at} &\qquad {\bf\small above}\\
    \hline
    \vspace{1mm}\\[-3mm]
    $\tau(Z)$
      & $0.07\pm 0.06$ & $2.18\pm 0.12$ & $1.86\pm 0.14$ \\
    \vspace{1mm}
    $\tau(\Nc)$
     & $0.02\pm0.01$ & $2.05\pm0.03$ & $0.03\pm0.01$ \\
    \hline
    \hline
  \end{tabular}
\caption{Scaling exponents $\alpha$ for the $N$-dependence $\tau\sim N^\alpha$ of the two
relaxation times $\tau(Z)$ and  $\tau(\Nc)$
in Fig.~\ref{relaxation_time} far below the transition ($F=-2.6\ll F^*
$ and $\varepsilon=-2.0 \ll \varepsilon^*$), at the transition
($F= F^*=0.00$ and $\varepsilon= \varepsilon^*=0.42$), and far above the
transition ($F=1.6\gg F^*$ and $\varepsilon=4.0 \gg \varepsilon^*$). }
\label{tab:notations2}
\end{table}

\begin{figure}[t]
\centering
\includegraphics[scale=0.34]{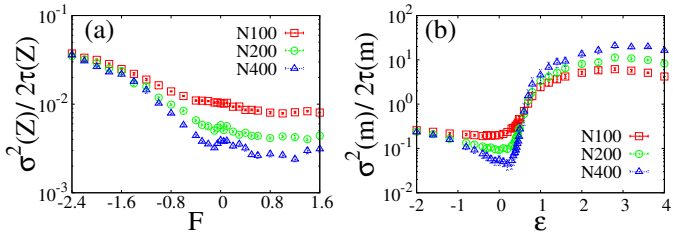}
\caption{(a) Effective diffusion coefficient
$D = \sigma^2(Z)/2 \tau(Z)$ of the chain end as a function
of applied force $F$. Panel (b) shows the corresponding quantity for $\Nc$,
which has the meaning of the effective frequency of monomer adsorption-desorption events.
}\label{diffuse}
\end{figure}

To quantify the global relaxation in the case of the loop-stretch transition,
one can introduce an effective longitudinal diffusion coefficient $D$, which
characterizes the motion of an ``equivalent" single particle in a harmonic
potential chosen such that the variance $\sigma^2(Z)$ is the same. From the
theory of Brownian motion, the effective diffusion constant can be estimated by
$D=\frac{\sigma^2(Z)}{2 \tau(Z)}$ \cite{Risken}.  The results for $D$ are
presented in Fig.~\ref{diffuse}(a).  Here, the variance $\sigma^2 (Z)$ is
obtained directly from the trajectory of the free end, and the relaxation time
$\tau (Z)$ is calculated from the autocorrelation functions.  Below the
loop-stretch transition, for negative forces $F$, $D$ decreases as the
transition is approached. In the transitional regime as well as for the
stretched coil, the diffusion coefficient roughly recovers the Rouse scaling
law for a free draining chain $D \sim N^{-1}$. Since the inverse diffusion
constant, $1/D$, corresponds to a friction, one can conclude that all chain
monomers contribute to the effective friction of stretched chains, whereas only
a fraction contributes for chains with ends pressed towards the surface. Over
all, however, the variations of $D$ are relatively small. The effective
diffusion constant does not change by orders of magnitude.

In the case of the adsorption transition, an analysis in terms of an effective
diffusion constant (and corresponding friction) is less straightforward.
However, one can imagine the relaxation process  as a sequence of more or less
correlated monomer adsorption/desorption events. Each event changes the number
of contacts by $\pm 1$. Taking this as an uncorrelated random walk along $\Nc$
axis with $t_0$ as the time per single step we obtain for the mean-square
displacement  as a function of time  $\langle (\Delta \Nc)^2\rangle(t)=t/t_0$.
Hence, the relaxation time scales as $\tau(\Nc) \sim t_0 \sigma^2(\Nc)$ , and
the ratio $ \sigma^2(\Nc)/ \tau(\Nc)$ has the meaning of the effective
frequency $t_0^{-1}$ of the adsorption-desorption steps. The fact that at
strong adsorption this frequency is proportional to $N$, see Fig.\
\ref{diffuse} (b),  means that the steps happen independently at different
locations along the chain and are, indeed, weakly correlated.  In the
near-critical region these steps appear to be  strongly correlated leading a
very low effective frequency $\sim N^{-1}$.  At strong repulsion, the frequency
increases again since the correlations are now limited to a fairly small part
of the chain adjacent to the grafting point. Overall the variation of frequency
$t_0^{-1}$ in the adsorption transition is much stronger than that of the
diffusion coefficient $D$ in the loop-stretch transition and shows a
non-monotonic behavior.

In the static case, we had established a formal analogy between the adsorption
and the loop-stretch transition in Sec.\ \ref{theory}. It was based on the
same form of the probability distributions for the order parameter at the critical point
leading to the crossover functions $\Psi_F$ and $\Psi_{\varepsilon}$
(Eqs.\ (\ref{eq:PsiF_2}) and (\ref{eq:PsiE_2})) of the same structure.
However, the underlying Hamiltonian does not
have this symmetry. The Hamiltonian of a strongly stretched end-grafted chain
at $\varepsilon=0$ can be approximated by that of an ideal free stretched chain; all
monomer-monomer correlations and their relaxation are determined by the normal
modes, and the relaxation mechanism is essentially diffusive.  The fact that
the relaxation time $\tau(Z)$ scales as $\tau \sim N^2$ for strongly
stretched chains indicates a typical Rouse behavior~\cite{Doi_Edwards}.  The
case of a strongly adsorbed chain at $F=0$ is quite different, since both the
attractive and the repulsive part of the surface interaction play essential
roles. Diagonalization of the Hamiltonian is not possible. Pair correlation
functions for monomers separated by a large contour distance (larger than the
adsorption blob size), are suppressed, and thus large-scale relaxations do not
contribute to time correlation function of the monomer-surface contact
number, $\langle m(0) m(t) \rangle$.
\begin{figure}[t]
   \includegraphics[scale=0.34]{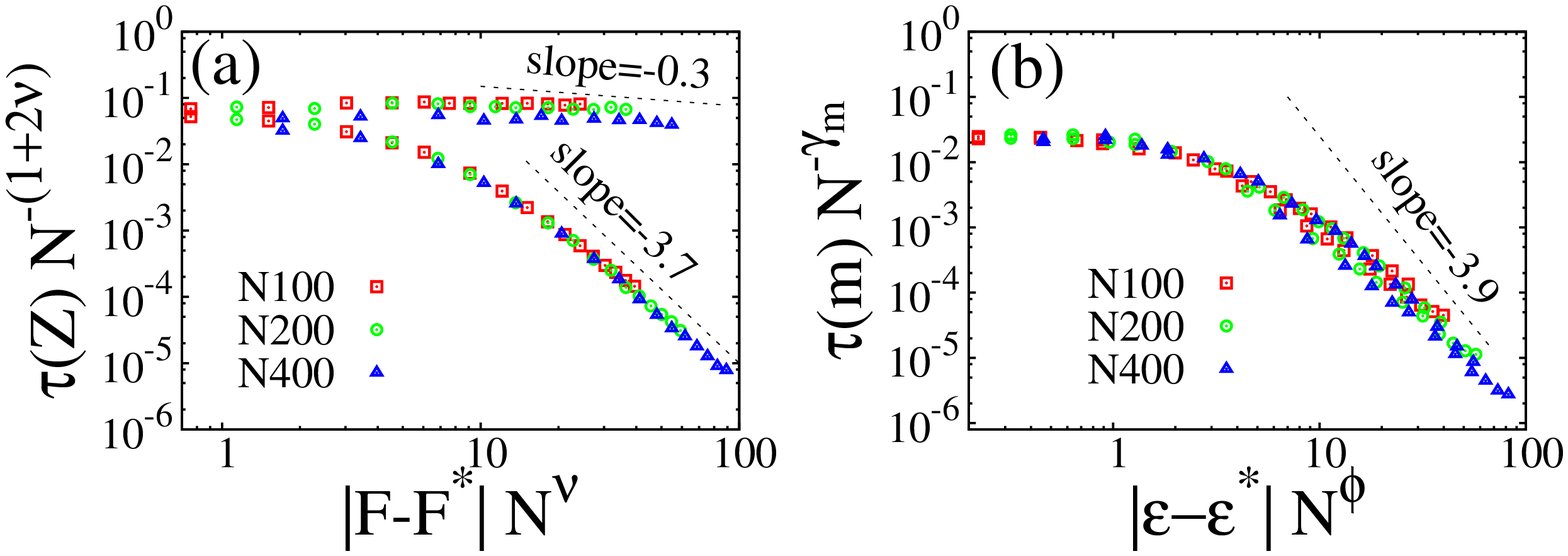}
  \caption{Same data as Fig.\ \protect\ref{relaxation_time} rescaled according to
Eq.\ (\protect\ref{eq:tau_z1}) with $\gamma_{_{Z}} \approx 1+ 2 \nu$ (a)
and Eq.\ (\protect\ref{eq:tau_nc}) with $\gamma_{_\Nc} = 2.05$ (b).
Lower branch in (a) and upper branch in (b) correspond to the regime
below the transition, upper branch in (a) and lower branch in (b)
to the regime above the transition.
}
\label{rescaled_tau}
\end{figure}

To further analyze the dynamic scaling behavior of the order parameters
of the loop-stretch and adsorption transition, we again make a scaling
Ansatz, similar to (\ref{eq:Fscaling}) and (\ref{eq:Escaling}).
\begin{eqnarray}
\label{eq:tau_z1}
\tau(Z)&=&N^{\gamma_{_{Z}}} \hat{f}_{\tau_{Z}}[(F-F^*)N^\nu]
\\
\label{eq:tau_nc}
\tau(\Nc)&=&N^{\gamma_{_\Nc}}
  \hat{f}_{\tau_{\Nc}} [(\varepsilon-\varepsilon^*)N^\phi]
\end{eqnarray}
For free chains, the characteristic relaxation time of the end-to-end distance
is expected to scale as $\tau \sim N^{1+2 \nu}$ (in the absence of
hydrodynamics) \cite{Doi_Edwards}. It seems reasonable to assume that this
exponent also sets the exponent $\gamma_{Z}$ in Eq.\ (\ref{eq:tau_z1}), i.e.,
one expects $\gamma_{Z} = 1+2\nu = 2.18$.  Indeed, this exponent is compatible
with the scaling of $\tau$ with chain length at the loop-stretch transition
(Table II), and Fig.\ \ref{rescaled_tau}(a) demonstrates that the data of Fig.\
\ref{relaxation_time}(a) for different chain lengths collapse nicely onto one
curve if they are rescaled as suggested by Eq.\ (\ref{eq:tau_z1}) with this
exponent.  For large negative $F$ (strongly pressed chain ends), the relaxation
of $Z$ is dominated by local processes close to the end monomer and expected to
be independent of the total chain length (see also Table II). Thus one expects
the scaling function $\hat{f}_{\tau_{Z}}$ to scale as $\hat{f}_{\tau_{Z}} (x)
\sim x^{-1/\nu - 2} \sim x^{-3.7}$, which is roughly compatible with the data
in Fig.\ \ref{rescaled_tau}(a) (lower branch).  For large positive $F$
(strongly stretched chain), the effective diffusion constant scales as $D \sim
1/N$ as discussed above, hence $\hat{f}_{\tau_{Z}}(x)$ must show the same
scaling behavior with $x$ than $\hat{f}_{_{\sigma^2(Z)}}(x)$,
$\hat{f}_{\tau_{Z}} \sim x^{1/\nu -2} \sim x^{-0.3}$, which is also compatible
with Fig.\ \ref{rescaled_tau}(a) (upper branch).

Since $\tau \sim N^{1+2 \nu}$ sets a characteristic time scale for grafted
chains in general, one might expect that it also characterizes the relaxation
time of the adsorption process \cite{MCAdsorption}, which would imply
$\gamma_{\Nc} = \gamma_{Z} = 2.18$.  However, the simulation data for
$\tau(\Nc)$ at different chain lengths do not collapse if they are rescaled
with that exponent (data not shown). Moreover, right at the transition,
$\tau(\Nc)$ was found to scale with $N$ with the exponent $2.05 \pm 0.03$
(see Table II), which suggests $\gamma_{\Nc} \approx 2.05$.
Rescaling the data with this exponent, one obtains good data collapse (see
Fig.\ \ref{rescaled_tau} (b)).  Far from the transition, the relaxation time
is found to be chain length independent both for $\varepsilon \ll
\varepsilon^*$ and $\varepsilon \gg \varepsilon^*$, which implies that the
scaling function $\hat{f}_{_{\tau(\Nc)}}$ should scale as
$N^{-\gamma_{\Nc}/\phi} \sim N^{-3.9}$. Fig.\ \ref{rescaled_tau} (b) indicates
that $\hat{f}_{_{\tau(\Nc)}}(x)$ indeed approaches this scaling for very large
arguments $|x|$.

It is clear that even within the error, $\gamma_m$ is still smaller than $1+2\nu$
which is expected to be the only characteristic time scale for grafted chains.
We do not have a good explanation for this observed difference. At the adsorption
critical point, the dynamics should be fully coupled, which precludes two
different global relaxation times with a different $N$ scaling. On the other
hand, Fig.\ \ref{relaxation_time} shows that the characteristic time scales
$\tau(Z)$ and $\tau(\Nc)$ are quite different even at the same state point
($F=0, \varepsilon=0$). Previous authors have also reported discrepancies
in the quality of the dynamic scaling of $\tau(Z)$ and
$\tau(\Nc)$ \cite{MCAdsorption}. It seems conceivable that the coupling
of the dynamics of the order parameters $Z$ and $\Nc$ is weak, such
that the slower relaxation time (with scaling $N^{\gamma_{_{Z}}}$) only
dominates in the limit of very long chains.

\section{Summary}\label{sec:summary}

To summarize, we have investigated the critical behavior of the
loop-stretch transition of end-grafted chains on neutral substrates by
Brownian Dynamics simulations and compared it with the critical behavior at the
adsorption transition of a loop. Loop adsorption was chosen to avoid possible
complications due to multicritical phenomena in the standard adsorption
transition of a single grafted chain.

We describe the loop-stretch transition in terms of the order parameter
$\langle Z\rangle$, the average height of the free end, and the adsorption
transition in terms of the order parameter $\langle \Nc \rangle$, the average
number of adsorbed segments. Both the analytical theory and the numerical
results suggest that the thermodynamic behavior of these two order
parameters is formally identical, with the same scaling with chain length, if
one replaces the crossover exponent $\phi$ in the adsorption transition by the
Flory exponent $\nu$ in the loop-stretch transition, and the crossover
variable $x=(\varepsilon - \varepsilon^*) N^\phi$ (in the adsorption
transition) by $x=(F - F^*) N^\nu$ (in the loop-stretch transition) (cf.\
Eqs.\ (\ref{eq:PsiF_2}) and (\ref{eq:PsiE_2})).

We show that both transitions provides a rare opportunity to
study in detail a continuous phase transition with the crossover behavior
characterized by two non-trivial critical exponents with an explicit analytical
form of the crossover function.  This exceptional
situation is made possible by a Fisher Ansatz  \cite{jcp_44_616} proposed for
distribution of the free end position
for a chain end-grafted to a non-adsorbing surface. Recognizing this as the order
parameter distribution  at the critical point of the loop-stretch transition we
extend the Ansatz to cover the loop adsorption transition as well.
In both cases, the order parameter distribution is characterized by a pair of indices
one of which is strictly defined by the relevant crossover exponent ($\nu$ or $\phi$) while the other
is also related to a pair of the partition function
surface exponents ($\gamma_1$, $\gamma_{11}$) or ($\gamma_{11c}$, $ \gamma_{11}$), respectively.
No fitting parameters are involved in the scaling form of the order parameter distribution function.
Explicit crossover function follows from Boltzmann re-weighting of the order parameter distribution
and provides an excellent agreement with the simulation results within the near-critical domain.

On the other hand, the dynamic relaxation behavior of the two order parameters
shows significant differences especially above the transition. While the
relaxation behavior for $Z$ above the loop-stretch transition shows a
typical Rouse behavior with relaxation times scaling as $\tau(Z) \sim N^2$, the
relaxation time for $m$ decays rapidly above the adsorption transition and
becomes essentially independent of the chain length far from the critical
point, i.e., the adsorption dynamics is local as in an Ising model. Right at
the critical point, critical slowing down is observed in both cases.  The
dynamic critical behavior at the adsorption transition is compatible with a
scaling Ansatz with crossover variable $x=(\varepsilon - \varepsilon^*)
N^\phi$, however, the scaling exponent $\gamma$ right at the transition
($\tau(m)|_{\varepsilon^*} \sim N^\gamma$) is smaller than the theoretically
expected value $\gamma = 1+2 \nu$. Likewise, the dynamic critical behavior at
the loop-stretch transition is described very well by a scaling Ansatz with
crossover variable $x = (F-F^*) N^\nu$ and in this case, the scaling exponent
takes the expected value, $\gamma = 1+2 \nu$.

To our best knowledge, the critical behavior of non-ideal chains at the
loop-stretch transition has been investigated for the first time in the present
work. Based on a conjecture by Fisher, we have derived an analytical expression
for the crossover function in the scaling hypothesis for the behavior at finite
$N$, which is in excellent quantitative agreement with our simulation data.
We have generalized the Fisher conjecture to be applied to the order parameter distribution
of the loop adsorption transition, and demonstrated that it provides an excellent
description of the near-critical scaling regime.
We believe that this approach could be further extended to cover continuous transitions
provided the order parameter is non-negative and interfacial phenomena do not interfere in the ordered state.
Coil-globule transition in a flexible chain could be a tentative candidate belonging to this class.

\bigskip
\begin{center}
\textbf{ACKNOWLEDGMENTS}
\end{center}

This work has been supported by the German Science Foundation (DFG) within the
Graduate School of Excellence Materials Science in Mainz (MAINZ), the grant
Schm 985/13-2, the grant NNIO-a 17-53-1213, and the SFB TRR 146 (project C1).  D.Y.  acknowledges financial
support from the National Natural Science Foundation of China (NSFC) 21374011, 21434001.
Simulations were carried out on the computer cluster Mogon at JGU
Mainz.

\begin{appendix}
\section{Brownian dynamics scheme}

In this Appendix we describe the simulation method in detail. We adopt the
over-damped Brownian dynamics to propagate the system (see Eq.(\ref{eq:BD})).
The system Hamiltonian is composed by two parts, i.e., $\mc H=\mc H_0+\mc
H_\mathrm I$, where $\mc H_0$ represents the contribution from the chain
connection, while $\mc H_\mathrm I$ is the interaction part. To be more general
we write the interaction part as a continuous integral
\begin{equation}
\mc H_\mathrm I=\frac{v}{2}\int d\mb r\hat\rho^2(\mb r)-\varepsilon\int d\mb rU_\mathrm a(\mb r)\hat\rho(\mb r)
\end{equation}
where the first term on the right hand side describes the excluded volume
interaction, while the second term is adsorption energy. The above Hamiltonian
corresponds to the system with pure adsorption, in the following we focus
mainly on the adsorption system. In the case of pure stretching, one could
directly add the stretching force to the first bead and remove the adsorption
force. The density operator at present can be defined as a delta function
$\hat\rho(\mb r)=\sum_j\delta(\mb r-\mb R_j)$, where the bead index $j$ runs
over all beads. By performing the derivative and using the chain rule, we
obtain
\begin{equation}
\frac{\partial\mc H}{\partial \mb R_j}=3(2\mb R_j-\mb R_{j+1}-\mb R_{j-1})+\frac{\partial}{\partial\mb R_j}\big[v\hat\rho-\varepsilon U_\mathrm a\big]
\end{equation}
where $j$ is valid for any intermediate beads. The first bead $j=1$ is only
connected to the second bead, while for the last bead $j=N$, it is always fixed
at the grafting point. By introducing a potential $\hat\omega\equiv
v\hat\rho-\varepsilon U_\mathrm a$, the BD equation for the first bead is
written as
\begin{equation}
\frac{d\mb R_{1}}{dt}=3(\mb R_2-\mb R_1)-\frac{\partial\hat\omega(\mb R_1)}{\partial\mb R_1}+\sqrt{2}\mb f_r,
\end{equation}
for the bead $j=N$, $\frac{d\mb R_{N}}{dt}=0$, for any bead $1<j<N$
\begin{equation}
\frac{d\mb R_{j}}{dt}=3(\mb R_\mathrm{j+1}+\mb R_\mathrm{j-1}-2\mb R_\mathrm{j})-\frac{\partial\hat\omega(\mb R_\mathrm{j})}{\partial\mb R_\mathrm{j}}+\sqrt{2}\mb f_r.
\end{equation}
In order to proceed, we need an explicit expression for the potential
derivative. Since the potential is directly determined by the bead density, the
evaluation of its derivative is coupled to the way of assignment of
particle-to-mesh density. In practice, we divide the simulation box uniformly
into $n_\mathrm x\cdot n_\mathrm y\cdot n_\mathrm z$ cells. All the quantities
are defined at the center of each cell, and we call these center points as mesh
points. Each cell has a volume of $\Delta V=l_\mathrm x\cdot l_\mathrm y\cdot
l_\mathrm z$. In the simulation we chose $l_\mathrm x=l_\mathrm y=l_\mathrm
z=a$, meaning that each cell has a unit volume.  Therefore the mesh points are
located at $x=0.5+ml_\mathrm x$, $y=0.5+nl_\mathrm y$, $z=0.5+ol_\mathrm z$,
where the integers $m\in[0,n_\mathrm x-1]$, $n\in[0,n_\mathrm y-1]$,
$o\in[0,n_\mathrm z-1]$.  Fractions of a bead are assigned to its neighbouring
mesh points according to the predefined assignment function $g(\mb r)$
depending only on the distance between the particle and mesh point. Rather than
choosing $g$ as a delta function in the continuum case, in practice $g$ has a
finite width playing the role of a smear function (or coarse-graining
function).

In terms of this assignment function, the density operator can be written as
$\hat\rho(\mb r_g)=\frac{1}{\Delta V}\sum_jg(|\mb R_j-\mb r_g|)$, where $\mb
r_g$ denotes the position of the $g$th mesh point. Now we can write $\mc
H_\mathrm I$ in a discretized form
\begin{equation}
\mc H_\mathrm I=\frac{v}{2}\Delta V\sum_g\hat\rho^2(\mb r_g)-\Delta V\varepsilon\sum_g U_\mathrm a(\mb r_g)\hat\rho(\mb r_g)
\end{equation}
which is equivalent to the corresponding interaction energy in Eq.(\ref{eq:HH})
as $\Delta V=1$. Then the derivative of $\mc H_\mathrm I$ can be performed to
get
\begin{equation}
\frac{\partial\mc H_\mathrm I}{\partial\mb R_j}=\frac{\partial\hat\omega}{\partial\mb R_j}=\sum_g\hat\omega(\mb r_\mathrm g)\frac{\partial}{\partial\mb R_j}g(|\mb R_j-\mb r_g|)
\end{equation}
To perform the derivative of the assignment function, we need its explicit
expression. For such a purpose, we consider the mesh in which $\mb R_j$ is
located. There are totally eight vertexes for the mesh, and let $i,j,k$ denote
the indices along $x,y,z$ directions, respectively. This means that $i=0, j=0,
k=0$ mark the vertex number 0 with coordinate $(0,0,0)$; $i=0, j=0, k=1$ is the
vertex number 1 with coordinate $(0,0,l_z)$, $i=0, j=1, k=0$ is the vertex
number 2 with coordinate $(0,l_y,0)$, and so on until $i=1,j=1,k=1$ is the the
vertex number 7 with coordinate $(l_x,l_y,l_z)$. Within this mesh, the $j$th
bead is located at $\mb R_j=(X,Y,Z)$. There are several choices for the
assignment function. The lowest order scheme is to assign each bead to its
nearest mesh point, and this is called the nearest-grid-scheme. Here we use a
higher order scheme, which assigns a fraction of bead to each of its eight
nearest mesh points. The fraction assigned to a given vertex is proportional to
the volume of a rectangle whose diagonal is the line connecting the particle
position and the mesh point on the opposite side of the mesh cell. With the
precise arrangement of vertexes, the assignment function for each vertex can be
written as
\begin{eqnarray}
\nonumber
\lefteqn{ g(\mb R_j-\mb r_g) } \nonumber \\
&&
=\frac{(l_x-|r_{gx}-X|)(l_y-|r_{gy}-Y|)(l_z-|r_{gz}-Z|)}{l_xl_yl_z}
\end{eqnarray}
where $g$ ranges from 0 to 7, $r_{g\alpha}$ is the $\alpha$ component of $\mb
r_g$. Performing the derivative of the assigning function directly, we obtain
\begin{eqnarray}
\frac{\partial\hat\omega}{\partial R_{jx}}&=&\frac{\hat \omega(\mb r_4)-\hat \omega(\mb r_0)}{l_x}\frac{(l_y-Y)(l_z-Z)}{l_yl_z}\nonumber\\
&+&\frac{\hat \omega(\mb r_5)-\hat \omega(\mb r_1)}{l_x}\frac{(l_y-Y)Z}{l_yl_z}\\
&+&\frac{\hat \omega(\mb r_6)-\hat \omega(\mb r_2)}{l_x}\frac{Y(l_z-Z)}{l_yl_z}+\frac{\hat \omega(\mb r_7)-\hat \omega(\mb r_3)}{l_x}\frac{YZ}{l_yl_z},\nonumber
\end{eqnarray}
\begin{eqnarray}
\frac{\partial\hat\omega}{\partial R_{jy}}&=&\frac{\hat \omega(\mb r_2)-\hat \omega(\mb r_0)}{l_y}\frac{(l_x-X)(l_z-Z)}{l_xl_z}\nonumber\\
&+&\frac{\hat \omega(\mb r_6)-\hat \omega(\mb r_4)}{l_y}\frac{X(l_z-Z)}{l_xl_z}\\
&+&\frac{\hat \omega(\mb r_3)-\hat \omega(\mb r_1)}{l_y}\frac{(l_x-X)Z}{l_xl_z}+\frac{\hat \omega(\mb r_7)-\hat \omega(\mb r_5)}{l_y}\frac{XZ}{l_xl_z},\nonumber
\end{eqnarray}and
\begin{eqnarray}
\frac{\partial\hat\omega}{\partial R_{jz}}&=&\frac{\hat \omega(\mb r_1)-\hat \omega(\mb r_0)}{l_z}\frac{(l_x-X)(l_y-Y)}{l_xl_y}\nonumber\\
&+&\frac{\hat \omega(\mb r_5)-\hat \omega(\mb r_4)}{l_z}\frac{X(l_y-Y)}{l_xl_y}\\
&+&\frac{\hat \omega(\mb r_3)-\hat \omega(\mb r_2)}{l_z}\frac{(l_x-X)Y}{l_xl_y}+\frac{\hat \omega(\mb r_7)-\hat \omega(\mb r_6)}{l_z}\frac{XY}{l_xl_y}.
\nonumber \end{eqnarray}
Inserting the above expressions to the BD equation gives the final form which
we use in our BD simulations.

We close with two remarks. First, a bead located at $z\le 0.5$ contributes
density only to the 4 nearest mesh points due to the impenetrable boundary
condition. This indicates that any bead at $z<0.5$ acts as if it were at
$z=0.5$.  Second, the adsorption potential is usually defined as a steplike
function with the potential width the segmental length $a$ ($a\equiv 1$ is the
unit length).  This means that $U_\mathrm a(z)=1$ for $z<1$ and zero otherwise.
In the present case, however, we introduce an assignment function to distribute
density to the nearest mesh points, which means that a bead can still feel the
force even it is at a location with $z>1$. Considering the specific form of the
assignment function, the apparent potential imposed on a bead should be
regulated as $U_\mathrm a(\mb r)=\min(1,3/2-z)$ for $z<3/2$, and zero
otherwise.

\end{appendix}


\begin{thebibliography}{45}%
\makeatletter
\providecommand \@ifxundefined [1]{%
 \@ifx{#1\undefined}
}%
\providecommand \@ifnum [1]{%
 \ifnum #1\expandafter \@firstoftwo
 \else \expandafter \@secondoftwo
 \fi
}%
\providecommand \@ifx [1]{%
 \ifx #1\expandafter \@firstoftwo
 \else \expandafter \@secondoftwo
 \fi
}%
\providecommand \natexlab [1]{#1}%
\providecommand \enquote  [1]{``#1''}%
\providecommand \bibnamefont  [1]{#1}%
\providecommand \bibfnamefont [1]{#1}%
\providecommand \citenamefont [1]{#1}%
\providecommand \href@noop [0]{\@secondoftwo}%
\providecommand \href [0]{\begingroup \@sanitize@url \@href}%
\providecommand \@href[1]{\@@startlink{#1}\@@href}%
\providecommand \@@href[1]{\endgroup#1\@@endlink}%
\providecommand \@sanitize@url [0]{\catcode `\\12\catcode `\$12\catcode
  `\&12\catcode `\#12\catcode `\^12\catcode `\_12\catcode `\%12\relax}%
\providecommand \@@startlink[1]{}%
\providecommand \@@endlink[0]{}%
\providecommand \url  [0]{\begingroup\@sanitize@url \@url }%
\providecommand \@url [1]{\endgroup\@href {#1}{\urlprefix }}%
\providecommand \urlprefix  [0]{URL }%
\providecommand \Eprint [0]{\href }%
\providecommand \doibase [0]{http://dx.doi.org/}%
\providecommand \selectlanguage [0]{\@gobble}%
\providecommand \bibinfo  [0]{\@secondoftwo}%
\providecommand \bibfield  [0]{\@secondoftwo}%
\providecommand \translation [1]{[#1]}%
\providecommand \BibitemOpen [0]{}%
\providecommand \bibitemStop [0]{}%
\providecommand \bibitemNoStop [0]{.\EOS\space}%
\providecommand \EOS [0]{\spacefactor3000\relax}%
\providecommand \BibitemShut  [1]{\csname bibitem#1\endcsname}%
\let\auto@bib@innerbib\@empty
\bibitem [{\citenamefont {Halperin}\ \emph {et~al.}(1992)\citenamefont
  {Halperin}, \citenamefont {Tirrell},\ and\ \citenamefont
  {Lodge}}]{advpolymsci.100.31}%
  \BibitemOpen
  \bibfield  {author} {\bibinfo {author} {\bibfnamefont {A.}~\bibnamefont
  {Halperin}}, \bibinfo {author} {\bibfnamefont {M.}~\bibnamefont {Tirrell}}, \
  and\ \bibinfo {author} {\bibfnamefont {T.~P.}\ \bibnamefont {Lodge}},\
  }\href@noop {} {\bibfield  {journal} {\bibinfo  {journal} {Adv. Polym. Sci.}\
  }\textbf {\bibinfo {volume} {100}},\ \bibinfo {pages} {31} (\bibinfo {year}
  {1992})}\BibitemShut {NoStop}%
\bibitem [{\citenamefont {Zhao}\ and\ \citenamefont
  {Brittain}(2000)}]{pps.25.677}%
  \BibitemOpen
  \bibfield  {author} {\bibinfo {author} {\bibfnamefont {B.}~\bibnamefont
  {Zhao}}\ and\ \bibinfo {author} {\bibfnamefont {W.~J.}\ \bibnamefont
  {Brittain}},\ }\href@noop {} {\bibfield  {journal} {\bibinfo  {journal}
  {Prog. Polym. Sci.}\ }\textbf {\bibinfo {volume} {25}},\ \bibinfo {pages}
  {677} (\bibinfo {year} {2000})}\BibitemShut {NoStop}%
\bibitem [{\citenamefont {Kato}\ \emph {et~al.}(2003)\citenamefont {Kato},
  \citenamefont {Uchida}, \citenamefont {Kang}, \citenamefont {Uyama},\ and\
  \citenamefont {Ikada}}]{pps.28.209}%
  \BibitemOpen
  \bibfield  {author} {\bibinfo {author} {\bibfnamefont {K.}~\bibnamefont
  {Kato}}, \bibinfo {author} {\bibfnamefont {E.}~\bibnamefont {Uchida}},
  \bibinfo {author} {\bibfnamefont {E.-T.}\ \bibnamefont {Kang}}, \bibinfo
  {author} {\bibfnamefont {Y.}~\bibnamefont {Uyama}}, \ and\ \bibinfo {author}
  {\bibfnamefont {Y.}~\bibnamefont {Ikada}},\ }\href@noop {} {\bibfield
  {journal} {\bibinfo  {journal} {Prog. Polym. Sci.}\ }\textbf {\bibinfo
  {volume} {28}},\ \bibinfo {pages} {209} (\bibinfo {year} {2003})}\BibitemShut
  {NoStop}%
\bibitem [{\citenamefont {Azzaroni}(2012)}]{jpsa.50.3225}%
  \BibitemOpen
  \bibfield  {author} {\bibinfo {author} {\bibfnamefont {O.}~\bibnamefont
  {Azzaroni}},\ }\href@noop {} {\bibfield  {journal} {\bibinfo  {journal} {J.
  Polym. Sci., Part A: Polym. Chem.}\ }\textbf {\bibinfo {volume} {50}},\
  \bibinfo {pages} {3225} (\bibinfo {year} {2012})}\BibitemShut {NoStop}%
\bibitem [{\citenamefont {Suo}\ and\ \citenamefont
  {Whitmore}(2014)}]{jcp.141.204903}%
  \BibitemOpen
  \bibfield  {author} {\bibinfo {author} {\bibfnamefont {T.}~\bibnamefont
  {Suo}}\ and\ \bibinfo {author} {\bibfnamefont {M.}~\bibnamefont {Whitmore}},\
  }\href@noop {} {\bibfield  {journal} {\bibinfo  {journal} {J. Chem. Phys.}\
  }\textbf {\bibinfo {volume} {141}},\ \bibinfo {pages} {204903} (\bibinfo
  {year} {2014})}\BibitemShut {NoStop}%
\bibitem [{\citenamefont {Qi}\ \emph {et~al.}(2015)\citenamefont {Qi},
  \citenamefont {Klushin}, \citenamefont {Skvortsov}, \citenamefont
  {Polotsky},\ and\ \citenamefont {Schmid}}]{ma2015}%
  \BibitemOpen
  \bibfield  {author} {\bibinfo {author} {\bibfnamefont {S.}~\bibnamefont
  {Qi}}, \bibinfo {author} {\bibfnamefont {L.~I.}\ \bibnamefont {Klushin}},
  \bibinfo {author} {\bibfnamefont {A.~M.}\ \bibnamefont {Skvortsov}}, \bibinfo
  {author} {\bibfnamefont {A.~A.}\ \bibnamefont {Polotsky}}, \ and\ \bibinfo
  {author} {\bibfnamefont {F.}~\bibnamefont {Schmid}},\ }\href@noop {}
  {\bibfield  {journal} {\bibinfo  {journal} {Macromolecules}\ }\textbf
  {\bibinfo {volume} {48}},\ \bibinfo {pages} {3775} (\bibinfo {year}
  {2015})}\BibitemShut {NoStop}%
\bibitem [{\citenamefont {Zhang}\ \emph {et~al.}(2017)\citenamefont {Zhang},
  \citenamefont {Qi}, \citenamefont {Klushin}, \citenamefont {Skvortsov},
  \citenamefont {Yan},\ and\ \citenamefont {Schmid}}]{Zhang_jcp}%
  \BibitemOpen
  \bibfield  {author} {\bibinfo {author} {\bibfnamefont {S.}~\bibnamefont
  {Zhang}}, \bibinfo {author} {\bibfnamefont {S.}~\bibnamefont {Qi}}, \bibinfo
  {author} {\bibfnamefont {L.~I.}\ \bibnamefont {Klushin}}, \bibinfo {author}
  {\bibfnamefont {A.~M.}\ \bibnamefont {Skvortsov}}, \bibinfo {author}
  {\bibfnamefont {D.}~\bibnamefont {Yan}}, \ and\ \bibinfo {author}
  {\bibfnamefont {F.}~\bibnamefont {Schmid}},\ }\href@noop {} {\bibfield
  {journal} {\bibinfo  {journal} {J. Chem. Phys}\ }\textbf {\bibinfo {volume}
  {147}},\ \bibinfo {pages} {064902} (\bibinfo {year} {2017})}\BibitemShut
  {NoStop}%
\bibitem [{\citenamefont {Urbakh}\ \emph {et~al.}(2004)\citenamefont {Urbakh},
  \citenamefont {Klafter}, \citenamefont {Gourdon},\ and\ \citenamefont
  {Israelachvili}}]{Adsorption}%
  \BibitemOpen
  \bibfield  {author} {\bibinfo {author} {\bibfnamefont {M.}~\bibnamefont
  {Urbakh}}, \bibinfo {author} {\bibfnamefont {J.}~\bibnamefont {Klafter}},
  \bibinfo {author} {\bibfnamefont {D.}~\bibnamefont {Gourdon}}, \ and\
  \bibinfo {author} {\bibfnamefont {J.}~\bibnamefont {Israelachvili}},\
  }\href@noop {} {\bibfield  {journal} {\bibinfo  {journal} {Nature}\ }\textbf
  {\bibinfo {volume} {430}},\ \bibinfo {pages} {525} (\bibinfo {year}
  {2004})}\BibitemShut {NoStop}%
\bibitem [{\citenamefont {Bustamante}\ \emph {et~al.}(1995)\citenamefont
  {Bustamante}, \citenamefont {Marko}, \citenamefont {Siggia},\ and\
  \citenamefont {Smith}}]{DNA1}%
  \BibitemOpen
  \bibfield  {author} {\bibinfo {author} {\bibfnamefont {C.}~\bibnamefont
  {Bustamante}}, \bibinfo {author} {\bibfnamefont {J.~F.}\ \bibnamefont
  {Marko}}, \bibinfo {author} {\bibfnamefont {E.~D.}\ \bibnamefont {Siggia}}, \
  and\ \bibinfo {author} {\bibfnamefont {S.}~\bibnamefont {Smith}},\
  }\href@noop {} {\bibfield  {journal} {\bibinfo  {journal} {Science}\ }\textbf
  {\bibinfo {volume} {265}},\ \bibinfo {pages} {1599} (\bibinfo {year}
  {1995})}\BibitemShut {NoStop}%
\bibitem [{\citenamefont {Smith}\ \emph {et~al.}(1996)\citenamefont {Smith},
  \citenamefont {Cui},\ and\ \citenamefont {Butamante}}]{DNA2}%
  \BibitemOpen
  \bibfield  {author} {\bibinfo {author} {\bibfnamefont {S.}~\bibnamefont
  {Smith}}, \bibinfo {author} {\bibfnamefont {Y.}~\bibnamefont {Cui}}, \ and\
  \bibinfo {author} {\bibfnamefont {C.}~\bibnamefont {Butamante}},\ }\href@noop
  {} {\bibfield  {journal} {\bibinfo  {journal} {Science}\ }\textbf {\bibinfo
  {volume} {271}},\ \bibinfo {pages} {795} (\bibinfo {year}
  {1996})}\BibitemShut {NoStop}%
\bibitem [{\citenamefont {Rief}\ \emph {et~al.}(1998)\citenamefont {Rief},
  \citenamefont {Fernandez},\ and\ \citenamefont {Gaub}}]{Protein}%
  \BibitemOpen
  \bibfield  {author} {\bibinfo {author} {\bibfnamefont {M.}~\bibnamefont
  {Rief}}, \bibinfo {author} {\bibfnamefont {J.~M.}\ \bibnamefont {Fernandez}},
  \ and\ \bibinfo {author} {\bibfnamefont {H.~E.}\ \bibnamefont {Gaub}},\
  }\href@noop {} {\bibfield  {journal} {\bibinfo  {journal} {Phys. Rev. Lett.}\
  }\textbf {\bibinfo {volume} {81}},\ \bibinfo {pages} {4764} (\bibinfo {year}
  {1998})}\BibitemShut {NoStop}%
\bibitem [{\citenamefont {Skvortsov}\ \emph {et~al.}(1994)\citenamefont
  {Skvortsov}, \citenamefont {Gorbunov},\ and\ \citenamefont
  {Klushin}}]{AdsorbedStretching}%
  \BibitemOpen
  \bibfield  {author} {\bibinfo {author} {\bibfnamefont {A.~M.}\ \bibnamefont
  {Skvortsov}}, \bibinfo {author} {\bibfnamefont {A.~A.}\ \bibnamefont
  {Gorbunov}}, \ and\ \bibinfo {author} {\bibfnamefont {L.~I.}\ \bibnamefont
  {Klushin}},\ }\href@noop {} {\bibfield  {journal} {\bibinfo  {journal} {J.
  Chem. Phys}\ }\textbf {\bibinfo {volume} {100}},\ \bibinfo {pages} {2325}
  (\bibinfo {year} {1994})}\BibitemShut {NoStop}%
\bibitem [{\citenamefont {Klushin}\ \emph {et~al.}(1997)\citenamefont
  {Klushin}, \citenamefont {Skvortsov},\ and\ \citenamefont
  {Gorbunov}}]{klushin97}%
  \BibitemOpen
  \bibfield  {author} {\bibinfo {author} {\bibfnamefont {L.~I.}\ \bibnamefont
  {Klushin}}, \bibinfo {author} {\bibfnamefont {A.~M.}\ \bibnamefont
  {Skvortsov}}, \ and\ \bibinfo {author} {\bibfnamefont {A.~A.}\ \bibnamefont
  {Gorbunov}},\ }\href@noop {} {\bibfield  {journal} {\bibinfo  {journal}
  {Phys. Rev. E}\ }\textbf {\bibinfo {volume} {56}},\ \bibinfo {pages} {1511}
  (\bibinfo {year} {1997})}\BibitemShut {NoStop}%
\bibitem [{\citenamefont {de~Gennes}(1979)}]{deGennes}%
  \BibitemOpen
  \bibfield  {author} {\bibinfo {author} {\bibfnamefont {P.~G.}\ \bibnamefont
  {de~Gennes}},\ }\href@noop {} {\emph {\bibinfo {title} {Scaling Concepts in
  Polymer Physics}}}\ (\bibinfo  {publisher} {Cornell University Press},\
  \bibinfo {year} {1979})\BibitemShut {NoStop}%
\bibitem [{\citenamefont {Eisenriegler}(1993)}]{Eisenriegler}%
  \BibitemOpen
  \bibfield  {author} {\bibinfo {author} {\bibfnamefont {E.}~\bibnamefont
  {Eisenriegler}},\ }\href@noop {} {\emph {\bibinfo {title} {Polymers Near
  Surfaces}}}\ (\bibinfo  {publisher} {World Scientific},\ \bibinfo {year}
  {1993})\BibitemShut {NoStop}%
\bibitem [{\citenamefont {Gorbunov}\ and\ \citenamefont
  {Skvortsov}(1993)}]{gorbunov93}%
  \BibitemOpen
  \bibfield  {author} {\bibinfo {author} {\bibfnamefont {A.~A.}\ \bibnamefont
  {Gorbunov}}\ and\ \bibinfo {author} {\bibfnamefont {A.~M.}\ \bibnamefont
  {Skvortsov}},\ }\href@noop {} {\bibfield  {journal} {\bibinfo  {journal} {J.
  Chem. Phys.}\ }\textbf {\bibinfo {volume} {98}},\ \bibinfo {pages} {5961}
  (\bibinfo {year} {1993})}\BibitemShut {NoStop}%
\bibitem [{\citenamefont {Skvortsov}\ \emph {et~al.}(2010)\citenamefont
  {Skvortsov}, \citenamefont {Klushin}, \citenamefont {Fleer},\ and\
  \citenamefont {Leermakers}}]{skvortsov10}%
  \BibitemOpen
  \bibfield  {author} {\bibinfo {author} {\bibfnamefont {A.~M.}\ \bibnamefont
  {Skvortsov}}, \bibinfo {author} {\bibfnamefont {L.~I.}\ \bibnamefont
  {Klushin}}, \bibinfo {author} {\bibfnamefont {G.~J.}\ \bibnamefont {Fleer}},
  \ and\ \bibinfo {author} {\bibfnamefont {F.~A.~M.}\ \bibnamefont
  {Leermakers}},\ }\href@noop {} {\bibfield  {journal} {\bibinfo  {journal} {J.
  Chem. Phys.}\ }\textbf {\bibinfo {volume} {132}},\ \bibinfo {pages} {064110}
  (\bibinfo {year} {2010})}\BibitemShut {NoStop}%
\bibitem [{\citenamefont {Klushin}\ and\ \citenamefont
  {Skvortsov}(2011)}]{klushin11}%
  \BibitemOpen
  \bibfield  {author} {\bibinfo {author} {\bibfnamefont {L.~I.}\ \bibnamefont
  {Klushin}}\ and\ \bibinfo {author} {\bibfnamefont {A.~M.}\ \bibnamefont
  {Skvortsov}},\ }\href@noop {} {\bibfield  {journal} {\bibinfo  {journal} {J.
  Phys. A}\ }\textbf {\bibinfo {volume} {44}},\ \bibinfo {pages} {473001}
  (\bibinfo {year} {2011})}\BibitemShut {NoStop}%
\bibitem [{\citenamefont {Skvortsov}\ \emph {et~al.}(2012)\citenamefont
  {Skvortsov}, \citenamefont {Klushin}, \citenamefont {Polotsky},\ and\
  \citenamefont {Binder}}]{skvortsov12}%
  \BibitemOpen
  \bibfield  {author} {\bibinfo {author} {\bibfnamefont {A.~M.}\ \bibnamefont
  {Skvortsov}}, \bibinfo {author} {\bibfnamefont {L.~I.}\ \bibnamefont
  {Klushin}}, \bibinfo {author} {\bibfnamefont {A.~A.}\ \bibnamefont
  {Polotsky}}, \ and\ \bibinfo {author} {\bibfnamefont {K.}~\bibnamefont
  {Binder}},\ }\href@noop {} {\bibfield  {journal} {\bibinfo  {journal} {Phys.
  Rev. E}\ }\textbf {\bibinfo {volume} {85}},\ \bibinfo {pages} {031803}
  (\bibinfo {year} {2012})}\BibitemShut {NoStop}%
\bibitem [{\citenamefont {Ioffe}\ and\ \citenamefont
  {Velenik}(2010)}]{BJPS_24_279}%
  \BibitemOpen
  \bibfield  {author} {\bibinfo {author} {\bibfnamefont {D.}~\bibnamefont
  {Ioffe}}\ and\ \bibinfo {author} {\bibfnamefont {Y.}~\bibnamefont
  {Velenik}},\ }\href@noop {} {\bibfield  {journal} {\bibinfo  {journal} {Braz.
  J. Probab. Stat.}\ }\textbf {\bibinfo {volume} {24}},\ \bibinfo {pages} {279}
  (\bibinfo {year} {2010})}\BibitemShut {NoStop}%
\bibitem [{\citenamefont {Beaton}(2015)}]{JPAMT_48_16FT03}%
  \BibitemOpen
  \bibfield  {author} {\bibinfo {author} {\bibfnamefont {N.~R.}\ \bibnamefont
  {Beaton}},\ }\href@noop {} {\bibfield  {journal} {\bibinfo  {journal} {J.
  Phys. A: Math. Theor.}\ }\textbf {\bibinfo {volume} {48}},\ \bibinfo {pages}
  {16FT03} (\bibinfo {year} {2015})}\BibitemShut {NoStop}%
\bibitem [{\citenamefont {van Rensburg}\ and\ \citenamefont
  {Whittington}(2013)}]{Whittington1}%
  \BibitemOpen
  \bibfield  {author} {\bibinfo {author} {\bibfnamefont {E.~J.~J.}\
  \bibnamefont {van Rensburg}}\ and\ \bibinfo {author} {\bibfnamefont {S.~G.}\
  \bibnamefont {Whittington}},\ }\href@noop {} {\bibfield  {journal} {\bibinfo
  {journal} {J. Phys. A: Math. Theor.}\ }\textbf {\bibinfo {volume} {46}},\
  \bibinfo {pages} {435003} (\bibinfo {year} {2013})}\BibitemShut {NoStop}%
\bibitem [{\citenamefont {van Rensburg}\ and\ \citenamefont
  {Whittington}(2016)}]{Whittington2}%
  \BibitemOpen
  \bibfield  {author} {\bibinfo {author} {\bibfnamefont {E.~J.~J.}\
  \bibnamefont {van Rensburg}}\ and\ \bibinfo {author} {\bibfnamefont {S.~G.}\
  \bibnamefont {Whittington}},\ }\href@noop {} {\bibfield  {journal} {\bibinfo
  {journal} {J. Phys. A: Math. Theor.}\ }\textbf {\bibinfo {volume} {49}},\
  \bibinfo {pages} {244001} (\bibinfo {year} {2016})}\BibitemShut {NoStop}%
\bibitem [{\citenamefont {Hammersley}\ \emph {et~al.}(1982)\citenamefont
  {Hammersley}, \citenamefont {Torrie},\ and\ \citenamefont
  {Whittington}}]{JPAMG_15_539}%
  \BibitemOpen
  \bibfield  {author} {\bibinfo {author} {\bibfnamefont {J.~M.}\ \bibnamefont
  {Hammersley}}, \bibinfo {author} {\bibfnamefont {G.~M.}\ \bibnamefont
  {Torrie}}, \ and\ \bibinfo {author} {\bibfnamefont {S.~G.}\ \bibnamefont
  {Whittington}},\ }\href@noop {} {\bibfield  {journal} {\bibinfo  {journal}
  {J. Phys. A: Math. Gen.}\ }\textbf {\bibinfo {volume} {15}},\ \bibinfo
  {pages} {539} (\bibinfo {year} {1982})}\BibitemShut {NoStop}%
\bibitem [{\citenamefont {Doi}\ and\ \citenamefont
  {Edwards}(1999)}]{Doi_Edwards}%
  \BibitemOpen
  \bibfield  {author} {\bibinfo {author} {\bibfnamefont {M.}~\bibnamefont
  {Doi}}\ and\ \bibinfo {author} {\bibfnamefont {S.~F.}\ \bibnamefont
  {Edwards}},\ }\href@noop {} {\emph {\bibinfo {title} {The theory of polymer
  dynamics}}}\ (\bibinfo  {publisher} {Oxford University Press},\ \bibinfo
  {year} {1999})\BibitemShut {NoStop}%
\bibitem [{\citenamefont {Hockney}\ and\ \citenamefont {Eastwood}(1988)}]{P3M}%
  \BibitemOpen
  \bibfield  {author} {\bibinfo {author} {\bibfnamefont {R.~W.}\ \bibnamefont
  {Hockney}}\ and\ \bibinfo {author} {\bibfnamefont {J.~W.}\ \bibnamefont
  {Eastwood}},\ }\href@noop {} {\emph {\bibinfo {title} {Computer Simulation
  using Particles}}}\ (\bibinfo  {publisher} {CRC Press},\ \bibinfo {year}
  {1988})\BibitemShut {NoStop}%
\bibitem [{\citenamefont {Detcheverry}\ \emph {et~al.}(2008)\citenamefont
  {Detcheverry}, \citenamefont {Kang}, \citenamefont {Daoulas}, \citenamefont
  {M\"{u}ller}, \citenamefont {Nealey},\ and\ \citenamefont {de~Pablo}}]{PtOM}%
  \BibitemOpen
  \bibfield  {author} {\bibinfo {author} {\bibfnamefont {F.~A.}\ \bibnamefont
  {Detcheverry}}, \bibinfo {author} {\bibfnamefont {H.}~\bibnamefont {Kang}},
  \bibinfo {author} {\bibfnamefont {K.~C.}\ \bibnamefont {Daoulas}}, \bibinfo
  {author} {\bibfnamefont {M.}~\bibnamefont {M\"{u}ller}}, \bibinfo {author}
  {\bibfnamefont {P.~F.}\ \bibnamefont {Nealey}}, \ and\ \bibinfo {author}
  {\bibfnamefont {J.~J.}\ \bibnamefont {de~Pablo}},\ }\href@noop {} {\bibfield
  {journal} {\bibinfo  {journal} {Macromolecules}\ }\textbf {\bibinfo {volume}
  {41}},\ \bibinfo {pages} {4989} (\bibinfo {year} {2008})}\BibitemShut
  {NoStop}%
\bibitem [{\citenamefont {Qi}\ \emph {et~al.}(2016)\citenamefont {Qi},
  \citenamefont {Klushin}, \citenamefont {Skvortsov},\ and\ \citenamefont
  {Schmid}}]{ma2016}%
  \BibitemOpen
  \bibfield  {author} {\bibinfo {author} {\bibfnamefont {S.}~\bibnamefont
  {Qi}}, \bibinfo {author} {\bibfnamefont {L.~I.}\ \bibnamefont {Klushin}},
  \bibinfo {author} {\bibfnamefont {A.~M.}\ \bibnamefont {Skvortsov}}, \ and\
  \bibinfo {author} {\bibfnamefont {F.}~\bibnamefont {Schmid}},\ }\href@noop {}
  {\bibfield  {journal} {\bibinfo  {journal} {Macromolecules}\ }\textbf
  {\bibinfo {volume} {49}},\ \bibinfo {pages} {9665} (\bibinfo {year}
  {2016})}\BibitemShut {NoStop}%
\bibitem [{\citenamefont {Birdsall}\ and\ \citenamefont {Fuss}(1997)}]{CIC}%
  \BibitemOpen
  \bibfield  {author} {\bibinfo {author} {\bibfnamefont {C.~K.}\ \bibnamefont
  {Birdsall}}\ and\ \bibinfo {author} {\bibfnamefont {D.}~\bibnamefont
  {Fuss}},\ }\href@noop {} {\bibfield  {journal} {\bibinfo  {journal} {J.
  Comput. Phys.}\ }\textbf {\bibinfo {volume} {135}},\ \bibinfo {pages} {141}
  (\bibinfo {year} {1997})}\BibitemShut {NoStop}%
\bibitem [{\citenamefont {Binder}\ and\ \citenamefont
  {Heermann}(2010)}]{Binder_cumulant}%
  \BibitemOpen
  \bibfield  {author} {\bibinfo {author} {\bibfnamefont {K.}~\bibnamefont
  {Binder}}\ and\ \bibinfo {author} {\bibfnamefont {D.~W.}\ \bibnamefont
  {Heermann}},\ }\href@noop {} {\emph {\bibinfo {title} {Monte Carlo Simulation
  in Statistical Physics: An Introduction}}}\ (\bibinfo  {publisher} {Springer
  Berlin Heidelberg},\ \bibinfo {year} {2010})\BibitemShut {NoStop}%
\bibitem [{\citenamefont {Clisby}\ and\ \citenamefont
  {D\"unweg}(2016)}]{Clisby1}%
  \BibitemOpen
  \bibfield  {author} {\bibinfo {author} {\bibfnamefont {N.}~\bibnamefont
  {Clisby}}\ and\ \bibinfo {author} {\bibfnamefont {B.}~\bibnamefont
  {D\"unweg}},\ }\href@noop {} {\bibfield  {journal} {\bibinfo  {journal}
  {Phys. Rev. E}\ }\textbf {\bibinfo {volume} {94}},\ \bibinfo {pages} {052102}
  (\bibinfo {year} {2016})}\BibitemShut {NoStop}%
\bibitem [{\citenamefont {Barber}\ \emph {et~al.}(1978)\citenamefont {Barber},
  \citenamefont {Guttmann}, \citenamefont {Middlemiss}, \citenamefont
  {Torrie},\ and\ \citenamefont {Whittington}}]{jpamg_11_1833}%
  \BibitemOpen
  \bibfield  {author} {\bibinfo {author} {\bibfnamefont {M.~N.}\ \bibnamefont
  {Barber}}, \bibinfo {author} {\bibfnamefont {A.~J.}\ \bibnamefont
  {Guttmann}}, \bibinfo {author} {\bibfnamefont {K.~M.}\ \bibnamefont
  {Middlemiss}}, \bibinfo {author} {\bibfnamefont {G.~M.}\ \bibnamefont
  {Torrie}}, \ and\ \bibinfo {author} {\bibfnamefont {S.~G.}\ \bibnamefont
  {Whittington}},\ }\href@noop {} {\bibfield  {journal} {\bibinfo  {journal}
  {J. Phys. A: Math. Gen.}\ }\textbf {\bibinfo {volume} {11}},\ \bibinfo
  {pages} {1833} (\bibinfo {year} {1978})}\BibitemShut {NoStop}%
\bibitem [{\citenamefont {Lax}(1974)}]{ma_7_660}%
  \BibitemOpen
  \bibfield  {author} {\bibinfo {author} {\bibfnamefont {M.}~\bibnamefont
  {Lax}},\ }\href@noop {} {\bibfield  {journal} {\bibinfo  {journal}
  {Macromolecules}\ }\textbf {\bibinfo {volume} {7}},\ \bibinfo {pages} {660}
  (\bibinfo {year} {1974})}\BibitemShut {NoStop}%
\bibitem [{\citenamefont {Mark}\ \emph {et~al.}(1975)\citenamefont {Mark},
  \citenamefont {Windwer},\ and\ \citenamefont {Lax}}]{ma_8_946}%
  \BibitemOpen
  \bibfield  {author} {\bibinfo {author} {\bibfnamefont {P.}~\bibnamefont
  {Mark}}, \bibinfo {author} {\bibfnamefont {S.}~\bibnamefont {Windwer}}, \
  and\ \bibinfo {author} {\bibfnamefont {M.}~\bibnamefont {Lax}},\ }\href@noop
  {} {\bibfield  {journal} {\bibinfo  {journal} {Macromolecules}\ }\textbf
  {\bibinfo {volume} {8}},\ \bibinfo {pages} {946} (\bibinfo {year}
  {1975})}\BibitemShut {NoStop}%
\bibitem [{\citenamefont {Ma}\ \emph {et~al.}(1977)\citenamefont {Ma},
  \citenamefont {Middlemiss},\ and\ \citenamefont {Whittington}}]{ma_10_1415}%
  \BibitemOpen
  \bibfield  {author} {\bibinfo {author} {\bibfnamefont {L.}~\bibnamefont
  {Ma}}, \bibinfo {author} {\bibfnamefont {K.~M.}\ \bibnamefont {Middlemiss}},
  \ and\ \bibinfo {author} {\bibfnamefont {S.~G.}\ \bibnamefont
  {Whittington}},\ }\href@noop {} {\bibfield  {journal} {\bibinfo  {journal}
  {Macromolecules}\ }\textbf {\bibinfo {volume} {10}},\ \bibinfo {pages} {1415}
  (\bibinfo {year} {1977})}\BibitemShut {NoStop}%
\bibitem [{\citenamefont {Bray}\ and\ \citenamefont
  {Moore}(1977)}]{jpamg_10_1927}%
  \BibitemOpen
  \bibfield  {author} {\bibinfo {author} {\bibfnamefont {A.~J.}\ \bibnamefont
  {Bray}}\ and\ \bibinfo {author} {\bibfnamefont {M.~A.}\ \bibnamefont
  {Moore}},\ }\href@noop {} {\bibfield  {journal} {\bibinfo  {journal} {J.
  Phys. A: Math. Gen.}\ }\textbf {\bibinfo {volume} {10}},\ \bibinfo {pages}
  {1927} (\bibinfo {year} {1977})}\BibitemShut {NoStop}%
\bibitem [{\citenamefont {Grassberger}(2005)}]{jpamg_38_323}%
  \BibitemOpen
  \bibfield  {author} {\bibinfo {author} {\bibfnamefont {P.}~\bibnamefont
  {Grassberger}},\ }\href@noop {} {\bibfield  {journal} {\bibinfo  {journal}
  {J. Phys. A: Math. Gen.}\ }\textbf {\bibinfo {volume} {38}},\ \bibinfo
  {pages} {323} (\bibinfo {year} {2005})}\BibitemShut {NoStop}%
\bibitem [{\citenamefont {Clisby}\ \emph {et~al.}(2016)\citenamefont {Clisby},
  \citenamefont {Conway},\ and\ \citenamefont {Guttmann}}]{Clisby2}%
  \BibitemOpen
  \bibfield  {author} {\bibinfo {author} {\bibfnamefont {N.}~\bibnamefont
  {Clisby}}, \bibinfo {author} {\bibfnamefont {A.~R.}\ \bibnamefont {Conway}},
  \ and\ \bibinfo {author} {\bibfnamefont {A.~J.}\ \bibnamefont {Guttmann}},\
  }\href@noop {} {\bibfield  {journal} {\bibinfo  {journal} {J. Phys. A: Math.
  Theor.}\ }\textbf {\bibinfo {volume} {49}},\ \bibinfo {pages} {015004}
  (\bibinfo {year} {2016})}\BibitemShut {NoStop}%
\bibitem [{\citenamefont {Clisby}(2017)}]{Clisby3}%
  \BibitemOpen
  \bibfield  {author} {\bibinfo {author} {\bibfnamefont {N.}~\bibnamefont
  {Clisby}},\ }\href@noop {} {\bibfield  {journal} {\bibinfo  {journal} {J.
  Phys. A: Math. Theor.}\ }\textbf {\bibinfo {volume} {50}},\ \bibinfo {pages}
  {264003} (\bibinfo {year} {2017})}\BibitemShut {NoStop}%
\bibitem [{\citenamefont {Fisher}(1966)}]{jcp_44_616}%
  \BibitemOpen
  \bibfield  {author} {\bibinfo {author} {\bibfnamefont {M.~E.}\ \bibnamefont
  {Fisher}},\ }\href@noop {} {\bibfield  {journal} {\bibinfo  {journal} {J.
  Chem. Phys.}\ }\textbf {\bibinfo {volume} {44}},\ \bibinfo {pages} {616}
  (\bibinfo {year} {1966})}\BibitemShut {NoStop}%
\bibitem [{\citenamefont {DiMarzio}(1965)}]{dimarzio_65}%
  \BibitemOpen
  \bibfield  {author} {\bibinfo {author} {\bibfnamefont {E.~A.}\ \bibnamefont
  {DiMarzio}},\ }\href@noop {} {\bibfield  {journal} {\bibinfo  {journal} {J.
  Chem. Phys}\ }\textbf {\bibinfo {volume} {42}},\ \bibinfo {pages} {2101}
  (\bibinfo {year} {1965})}\BibitemShut {NoStop}%
\bibitem [{\citenamefont {Diehl}\ and\ \citenamefont {Shpot}(1998)}]{diehl98}%
  \BibitemOpen
  \bibfield  {author} {\bibinfo {author} {\bibfnamefont {H.-W.}\ \bibnamefont
  {Diehl}}\ and\ \bibinfo {author} {\bibfnamefont {M.}~\bibnamefont {Shpot}},\
  }\href@noop {} {\bibfield  {journal} {\bibinfo  {journal} {Nuc. Phys. B}\
  }\textbf {\bibinfo {volume} {528}},\ \bibinfo {pages} {595} (\bibinfo {year}
  {1998})}\BibitemShut {NoStop}%
\bibitem [{\citenamefont {Descas}\ \emph {et~al.}(2004)\citenamefont {Descas},
  \citenamefont {Sommer},\ and\ \citenamefont {Blumen}}]{MCAdsorption}%
  \BibitemOpen
  \bibfield  {author} {\bibinfo {author} {\bibfnamefont {R.}~\bibnamefont
  {Descas}}, \bibinfo {author} {\bibfnamefont {J.-U.}\ \bibnamefont {Sommer}},
  \ and\ \bibinfo {author} {\bibfnamefont {A.}~\bibnamefont {Blumen}},\
  }\href@noop {} {\bibfield  {journal} {\bibinfo  {journal} {J. Chem. Phys.}\
  }\textbf {\bibinfo {volume} {120}},\ \bibinfo {pages} {8831} (\bibinfo {year}
  {2004})}\BibitemShut {NoStop}%
\bibitem [{\citenamefont {Klushin}\ \emph {et~al.}(2013)\citenamefont
  {Klushin}, \citenamefont {Polotsky}, \citenamefont {Hsu}, \citenamefont
  {Markelov}, \citenamefont {Binder},\ and\ \citenamefont
  {Skvortsov}}]{klushin13}%
  \BibitemOpen
  \bibfield  {author} {\bibinfo {author} {\bibfnamefont {L.~I.}\ \bibnamefont
  {Klushin}}, \bibinfo {author} {\bibfnamefont {A.~A.}\ \bibnamefont
  {Polotsky}}, \bibinfo {author} {\bibfnamefont {H.-P.}\ \bibnamefont {Hsu}},
  \bibinfo {author} {\bibfnamefont {D.~A.}\ \bibnamefont {Markelov}}, \bibinfo
  {author} {\bibfnamefont {K.}~\bibnamefont {Binder}}, \ and\ \bibinfo {author}
  {\bibfnamefont {A.~M.}\ \bibnamefont {Skvortsov}},\ }\href@noop {} {\bibfield
   {journal} {\bibinfo  {journal} {Phys. Rev. E}\ }\textbf {\bibinfo {volume}
  {87}},\ \bibinfo {pages} {022604} (\bibinfo {year} {2013})}\BibitemShut
  {NoStop}%
\bibitem [{\citenamefont {Risken}(1996)}]{Risken}%
  \BibitemOpen
  \bibfield  {author} {\bibinfo {author} {\bibfnamefont {H.}~\bibnamefont
  {Risken}},\ }\href@noop {} {\emph {\bibinfo {title} {The Fokker-Planck
  Equation}}},\ \bibinfo {series} {Sprinter Series in Synergetics},
  Vol.~\bibinfo {volume} {18}\ (\bibinfo  {publisher} {Springer},\ \bibinfo
  {year} {1996})\BibitemShut {NoStop}%
\end{thebibliography}
\end{document}